\documentclass[10pt,journal,cspaper,compsoc]{IEEEtran}
\ifCLASSOPTIONcompsoc
\else
\fi

\ifCLASSINFOpdf
\else
\fi

\ifCLASSINFOpdf
   \usepackage[pdftex]{graphicx}
\else

\fi

\usepackage{cite}
\usepackage{url,epsfig,times,amsmath}
\usepackage{array}
\usepackage{mdwmath}
\usepackage{mathtools}
\usepackage{multirow}
\usepackage[outdir=./Figs/]{epstopdf}
\usepackage[tight,footnotesize]{subfigure}

\begin{document}
\title{Performance Analysis of CSMA/CA based Medium Access in Full Duplex Wireless Communications}

\author{Rahman Doost-Mohammady,~\IEEEmembership{Member,~IEEE,}
		M. Yousof Naderi,~\IEEEmembership{Student Member,~IEEE,}
		and~Kaushik Roy Chowdhury,~\IEEEmembership{Senior Member,~IEEE}% <-this % stops a space
\IEEEcompsocitemizethanks{\IEEEcompsocthanksitem R. Doost-Mohammady, M.Y. Naderi and K.R. Chowdhury are with the Department
	of Electrical and Computer Engineering, Northeastern University, Boston,
	MA, 02115 USA e-mail: doost,naderi,krc@ece.neu.edu}% <-this % stops a space
\thanks{Manuscript received April 20, 2014; revised December 2 2014, May 4 and July 17, 2015; Accepted July 20, 2015. Citation
information: DOI 10.1109/TMC.2015.2462832, IEEE Transactions on Mobile Computing}}

\markboth{This article has been accepted for publication in a future issue of IEEE Transactions on Mobile Computing. }%
{Doost-Mohammady \MakeLowercase{\textit{et al.}}: Performance Analysis of CSMA/CA based Medium Access in Full Duplex Wireless Communications}

% PRIOR to the title within the \IEEEcompsoctitleabstractindextext IEEEtran
\IEEEcompsoctitleabstractindextext{%
\begin{abstract}
%\boldmath
Full duplex communication promises a paradigm shift in wireless networks by allowing simultaneous packet transmission and reception within the same channel. While recent prototypes indicate the feasibility of this concept, there is a lack of rigorous theoretical development on how full duplex impacts medium access control (MAC) protocols in practical wireless networks. In this paper, we formulate the first analytical model of a CSMA/CA based full duplex MAC protocol for a wireless LAN network composed of an access point serving mobile clients. There are two major contributions of our work: First, our Markov chain-based approach results in closed form expressions of throughput for both the access point and the clients for this new class of networks. Second, our study provides quantitative insights on how much of the classical hidden terminal problem can be mitigated through full duplex. We specifically demonstrate that the improvement in the network throughput is up to 35-40\% over the half duplex case. Our analytical models are verified through packet level simulations in ns-2. Our results also reveal the benefit of full duplex under varying network configuration parameters, such as number of hidden terminals, client density, and contention window size.
\end{abstract}

% Note that keywords are not normally used for peer review papers.
}

\maketitle

% is not selected <OR> if conference mode is selected - because compsoc
% conference papers position the abstract like regular (non-compsoc)
\IEEEdisplaynotcompsoctitleabstractindextext

% \fi
\IEEEpeerreviewmaketitle

\section{Introduction}

\IEEEPARstart{W}{ireless} full duplex (FD) technology allows a radio to send and receive data on the same channel simultaneously. It promises massive improvements in channel capacity by ushering in a paradigm shift in the design of existing networking protocols~\cite{MSRFDReport09,MSRFDReport27,Choi10,Sahai11,Duarte10,Duarte11,Duarte12}. Before its inception, half duplex communication was the de-facto standard, i.e., nodes may either transmit or receive at any given time~\cite{Goldsmith}. This key assumption influenced the design of the protocol stack, especially the channel access mechanism at the link layer. As a result, any simultaneous use of the channel by more than one node within their interference range in the same network could cause the transmitted packets to collide. This, in turn, results in wastage of bandwidth resources, and brings in the requirement for retransmission by all of the contending nodes that suffered packet losses. With the advent of the FD technology and the ability to 
transmit and receive at the same time on the wireless channel, this problem of simultaneous channel access can be mitigated to some extent. While recent work on building such systems~\cite{Jain11,Bharadia13} are important steps towards practical realizations of this technology, there has been very limited work on analyzing FD performance for protocol stack implementations. This paper attempts to bridge this gap at the link layer by assuming a simple CSMA/CA channel access, and then defining a Markov chain-based theoretical model to give closed form expressions for the system performance. 

\begin{figure}[!t]
\centering
\includegraphics[width=0.5\textwidth]{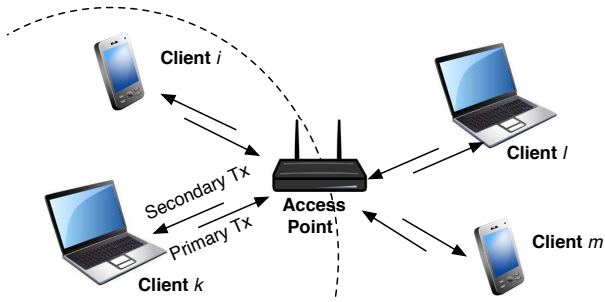}
\caption{Representation of a Star topology with full duplex nodes.}
\label{fig:net}
\end{figure}

%XXX Draw a small network diagram, with the AP i at the center and few nodes around, include j and k. Refer to that diagram in the following
For the analysis presented in this paper, we formally define the transmission scenario shown in Fig.~\ref{fig:net}, wherein an access point (AP) has several associated clients, with each device equipped with a FD radio. Using terminology similar to~\cite{MSRFDReport27}, let client $k$ be the \textit{primary transmitter} that sends its own packet to the AP, which now assumes the role of the \textit{secondary transmitter}. The secondary transmitter, upon receiving the primary transmitter's packet, can potentially start transmitting its own packet, thereby creating a \textit{dual} link.
%The secondary receiver $k$ may also be a node not in the range of the primary transmitter $i \ne k$ so that an \textit{dual asymmetric} link is formed. 
Client $i$ that lies within the coverage radius of client $k$ immediately detects the transmission and postpones its own transmissions if any. Additionally, by letting the AP to transmit while it is receiving, other clients hidden from $k$, namely $l$ and $m$ detect the channel as busy, and refrain from transmitting any packets themselves. Thus, the FD channel access avoids collisions and mitigates the hidden terminal problem to a considerable extent.

Despite its benefits, FD brings in several unique challenges in MAC protocol design. To realize its full potential, the intended receiver, (AP in Fig.~\ref{fig:net}, \textit{must} have packets for the primary transmitter $k$ at the same time. This can be determined by examining the header of the incoming packet (before the packet has completely been delivered to save on processing time) to determine the transmitter, and checking if the head-of-line (HOL) packet at the receiver is in fact addressed to the primary transmitter. Moreover, there are fairness concerns regarding channel access, as the AP \textit{piggybacks} on the original contention resolution won by client $k$.  While this improves the overall channel utilization, other nodes may potentially sacrifice their own access time to allow the AP to continue its transmissions. %Thus, efforts on MAC scheduling for full duplex attempt to schedule as many duplex links as possible, to get the maximum benefit. 
In an attempt to mitigate this problem of fair channel access, existing works on FD ~\cite{MSRFDReport27,Sahai11,Jain11} modify classical CSMA/CA and 802.11 DCF, where an adaptive back off counter is used for randomizing channel access more fairly among nodes. Although these works have evaluated the performance of their MAC through simulation and implementation, there is a lack of a rigorous analytical model quantifying their operational benefits. While there are numerous seminal works on the analysis of MAC protocols for classical half duplex networks~\cite{bianchi,malone07,Wu06,Tsertou08,Hung08,Hung10,Dai13,Gao13,Jang12}, they do not reveal insights on the performance gain of FD over classical half duplex (HD) scenarios. Moreover, the impact of various network settings, including the number of nodes, their traffic rates, number of hidden terminals, and the selection of the backoff duration cannot be obtained through trivial extensions of these half duplex models. For these reasons, we formulate a completely new theoretical model specially suited for FD in this paper.

 The main contributions of our work are as follows:
 \begin{itemize}
 \item This is the first work that provides a rigorous theoretical framework for analyzing FD networks using CSMA/CA as the channel access mechanism at the link layer. We consider a general scenario of an AP controlled wireless LAN, with the AP at the center, surrounded by the serviced clients in a \textit{star} topology. 
 \item  We derive closed form expressions for the probability of successful transmission and throughput separately for the AP and clients, while considering the effect of hidden terminals on these performance metrics.  

% by relaxing the heterogeneity among nodes. We assume homogeneous packet arrival and the possibility of hidden terminals for all the client nodes. We maintain separate formulations for the AP and clients, since the AP has no hidden terminals and different packet arrival than the clients. 
% \item We assume all the nodes in the network are in saturated conditions meaning they always have packets to transmit in their queue. This will lead to understanding how full duplex operates under realistic, high traffic conditions.
 \item Apart from considering FD, our work contributes to the existing works on MAC layer analysis in the presence of hidden terminals by separately analyzing both uplink traffic flows from clients to the AP, and downlink traffic from the AP to client. This is in contrast to ~\cite{Hung10,Tsertou08,Wu06,Liu10} where only uplink traffic is considered, and the AP is assumed to be only receiving. The consideration of both uplink and downlink traffic is imperative because FD is based on active bi-directional links.
\end{itemize}

Our theoretical findings are verified through packet level simulations in ns-2, which has been considerably modified from its stock installation to incorporate the FD operation.% and additional features such as busy tone, all will be explained in the next section. 

The rest of this paper is organized as follows: In Section~\ref{sec:related} we review related work, and in Section~\ref{sec:arch}, we discuss the network architecture. In Section~\ref{sec:model}, we give a mathematical model of the FD MAC protocol and derive closed-form expressions. In Section~\ref{sec:validation}, we validate our model with extensive simulation results and conclude the paper in Section~\ref{sec:conc}.

%In this section, we review the very few existing MAC algorithms for full duplex. The work of Radunovic et al. \cite{MSRFDReport27} was the first of its kind where a CSMA/CA based algorithm was used to schedule channel access for nodes. The node winning the contention will initiate its transmission and 

\section{Related Work and Motivation}
\label{sec:related}
%XXX Move the following discussion on routing to "related work"
%Full duplex feature in wireless nodes, as useful as it is, might not be used at all times due to various issues, for e.g. unavailability of a packet at the primary receiver. In the case of mesh networks, routing algorithms could be revised to maximize the usage of full duplex by directing the traffic in such that as many symmetric and asymmetric links as possible are created. A few works like~\cite{Fang11} study this aspect of full duplex.

The design of efficient protocol stack for FD networks is in an early stage. In this section, we review the existing works for MAC protocol design and evaluation of FD networks. 
A full–duplex MAC protocol, called ContraFlow, is proposed in \cite{Singh11} along with the development of a prototype that includes a back-off algorithm for improving fairness. The performance evaluation is limited to networks of limited size and selected topologies. 
In~\cite{Sahai11}, the authors proposed a distributed full-duplex MAC, called FD-MAC, which introduces features such as shared random back-off and virtual contention resolution with their corresponding implementations on the WARP platform~\cite{warp}. In FD-MAC, AP switches between full duplex and half duplex to ensure that all nodes get a chance to transmit. 
A MAC with dynamic contention window control based on the current transmission queue length is proposed in~\cite{Oashi12} to increase the transmission opportunity of FD operation, and to balance uplink and downlink traffic. However, simulation results are limited to sparse topologies, without including scenarios involving hidden terminals.
In~\cite{Jain11}, a simple CSMA/CA based MAC protocol is implemented on the WARP platform that broadcasts a busy tone (BT) by the AP to eliminate the hidden terminal problem during an empty slot.
In~\cite{Zhou13}, a distributed FD MAC protocol for ad-hoc networks and multi-AP networks is proposed in order to maximize FD and concurrent transmissions in the network, using additional signaling based on pseudo-random noise sequences.
Janus~\cite{kim2013janus} is another full-duplex MAC protocol that is centralized at the AP, and eliminates random back-off. AP collects interference information from nodes, divides transmission schedules based on the global traffic information, and send control packets at the beginning and end of each round to avoid collisions.

The above works laid the initial foundations of how protocols and hardware that support FD operation may be designed. However, they do not include an analytical framework that can be used to predict the performance of FD networks in general network settings, such as varying number of nodes and hidden terminals, contention window length, among others.

There has been some recent effort in characterizing FD's performance from a theoretical standpoint.
In \cite{Yang14}, achievable throughput in full-duplex is characterized as opposed to other channel access schemes such as MIMO and MU-MIMO. % if there is anything significant there, mention it!
In \cite{Xie14}, theoretical bounds for full-duplex gain over half-duplex has been derived for various topologies as a function of difference between transmission and interference range. It has been shown that when these two ranges are equal for a randomly deployed ad hoc network the asymptotic bound for full-duplex gain is only $28 \%$. 
However, none of these works consider a methematical modeling of a real-world FD protocol. Our work serves in bridging this gap, and we use the CSMA/CA based MAC protocol in~\cite{Jain11} implemented on physical hardware, as the base protocol with few additional modifications to its busy tone broadcasting scenarios. 
In this work, we extend the model in~\cite{Dai13}, which is an accurate analytical model of a saturated IEEE 802.11 DCF network with no hidden terminals, to the full-duplex medium access network with the presence of hidden terminals. 
\section{Network Architecture}
\label{sec:arch}
Consider the network shown in Fig.~\ref{fig:net}. Let $n$ clients be connected to an AP. Each node $i$ has a set of {\em covered} nodes $\mathcal{N}_c^i$ that can hear its transmission, and a set of hidden nodes $\mathcal{N}_h^i$ which are out of its sensing range (assume equal sensing and receiving range for all nodes). Each node, including the AP, adopts a CSMA/CA channel access mechanism with a contention window (CCW). More specifically, to access the medium, each node with a packet to transmit randomly chooses a value in the range $[0,W)$ and counts down to zero from that value during the time it senses the channel as idle. This means that if the channel gets busy during countdown, the back off timer is frozen. The countdown is resumed once the channel becomes idle again, and once this number reaches zero, the node attempts transmission. 

\begin{figure}
\centering
\subfigure[\label{fig:FD1}]{\includegraphics[width=0.4\textwidth]{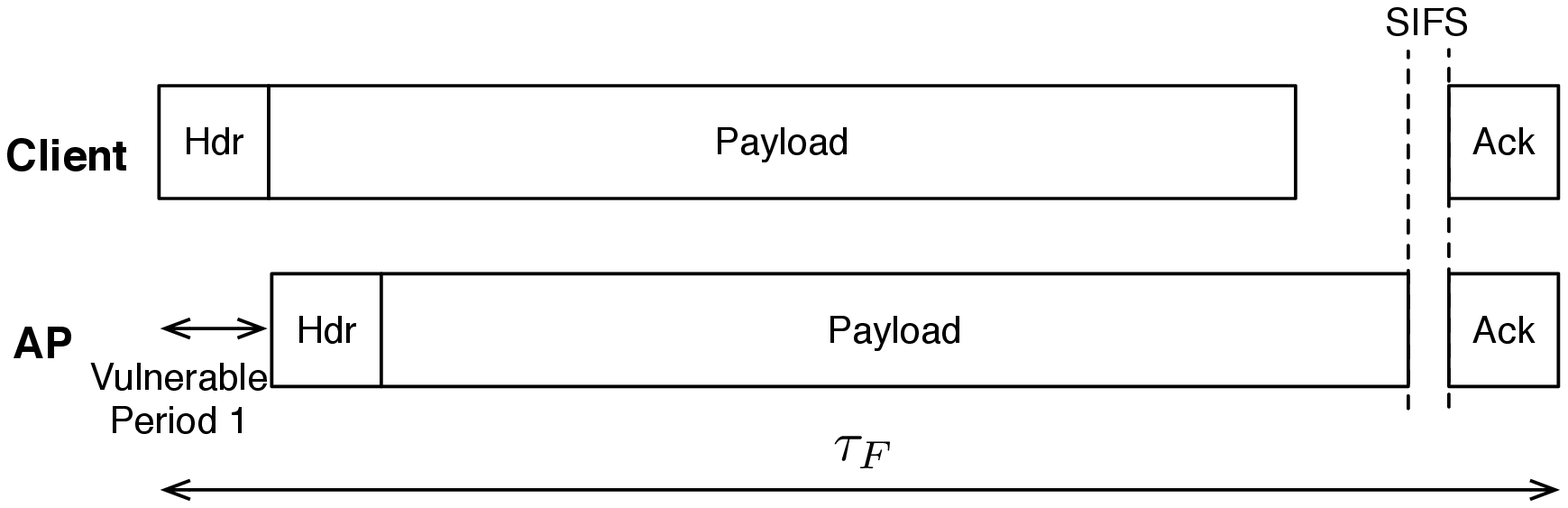}} 
\subfigure[\label{fig:HD2}]{\includegraphics[width=0.4\textwidth]{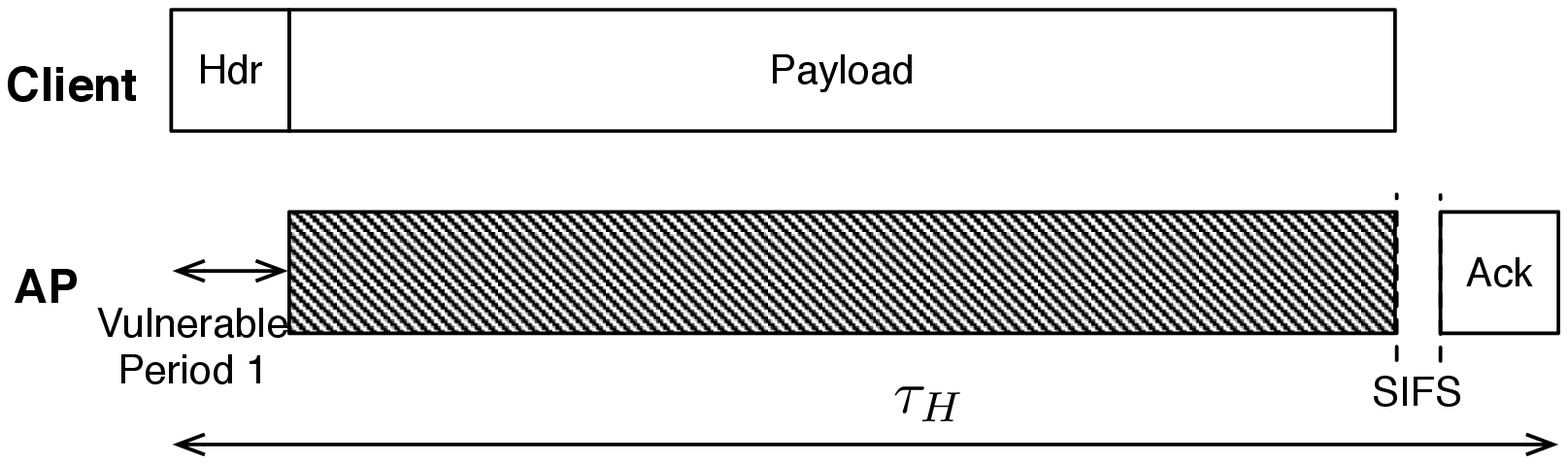}} 
\subfigure[\label{fig:FD2}]{\includegraphics[width=0.4\textwidth]{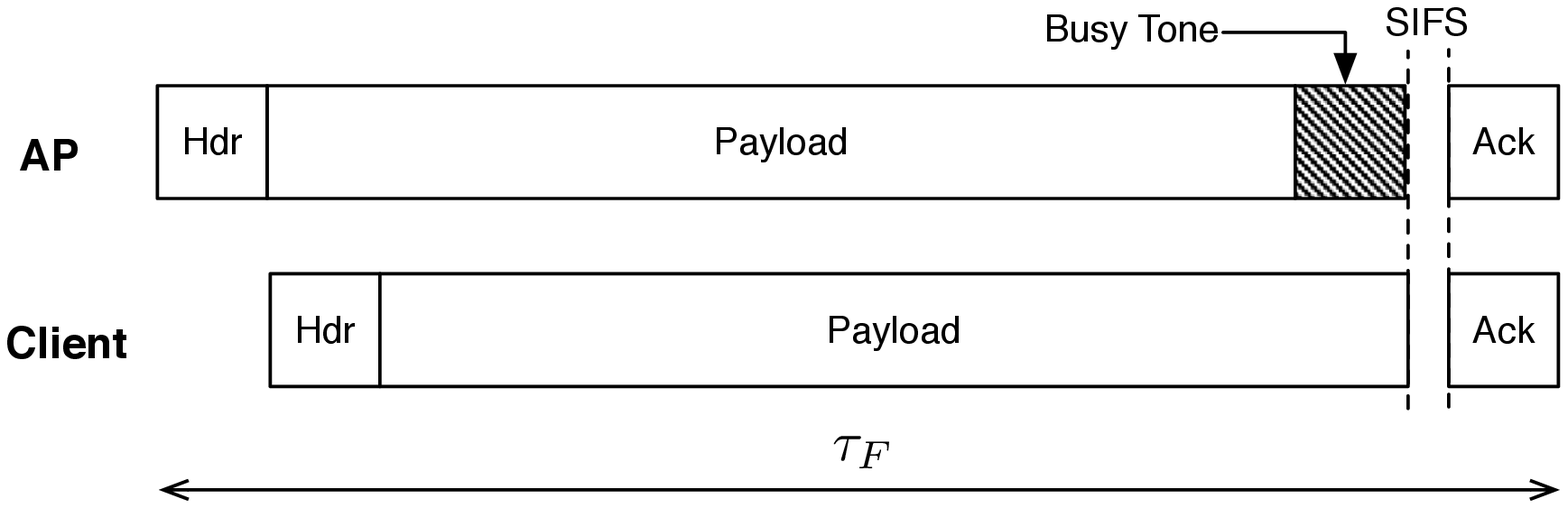}} 
\caption{Cases of full duplex transmissions initiated by (a) the client node and a packet reply by the AP, (b) the client node and BT broadcast by the AP, and (c) the AP and packet reply by the client.}
\label{fig:FD}
\end{figure}

We assume all nodes have FD capability, and hence, once a given node initiates transmission to a destination node, the latter checks whether it has a packet for the former at the HOL. If the HOL packet is destined for the source node, the destination starts sending it in FD mode to the source. Fig.~\ref{fig:FD1} shows such a scenario, where the client initiates packet transmission to the AP. The latter decodes the packet header, and compares the sending node's address to the destination of its own HOL packet. If {\em they are the same}, it enters FD mode and sends the HOL packet. If {\em not}, the AP sends a BT in order to keep the channel busy and prevent any hidden terminal of the client from transmitting and causing a collision. This scenario is shown in Fig.~\ref{fig:HD2}. In this case, the length of transmission and channel busy time is $\tau_H$ time slots. 

Fig.~\ref{fig:FD2} shows the AP initiating a packet transmission to a client. Since all clients are sending only to the AP, the HOL packet at the client is addressed to the AP by default. Therefore, the client enters FD mode and immediately sends its packet as it begins to receive a packet from the AP. Assuming fixed packet length for both nodes engaging in the FD data exchange, the AP notifies the hidden terminals by broadcasting a BT at the end of its packet transmission, while the client is still transmitting. At the end of a successful FD transmission, the two nodes send their ACKs simultaneously to each other after a fixed short gap called SIFS, as per the standard CSMA/CA algorithm (the processing time needed to check the correctness of the received packet). The total number of slot times for a successful FD transmission is assumed $\tau_F$ time slots in both Figs.~\ref{fig:FD1} and~\ref{fig:FD2}.

%\begin{figure}
%\centering
%\subfigure[\label{fig:HD1}]{\includegraphics[width=0.4\textwidth]{Figs/HD1}} 
%\subfigure[\label{fig:HD2}]{\includegraphics[width=0.4\textwidth]{Figs/HD2}} 
%\caption{half duplex  is initiated by (a) client node (b) AP.}
%\label{fig:HD}
%\end{figure}
%
The combination of FD and BT does not fully eliminate the hidden terminal problem as the header transmission of the primary transmitter is susceptible to collision with some probability. This period is shown in Figs.~\ref{fig:FD1} and~\ref{fig:HD2} and called \textit{vulnerable period}. We assume fixed header time for all the nodes in our system, consuming $\tau_{V}$ time slots.

% discuss the detailts of collision length
As we discussed earlier, FD does not fully remove the hidden terminal problem and collisions might still occur. Furthermore, two or more nodes, hidden or covered, might start transmission at the same slot hence causing a collision at the receiver. In all these cases the length of collision will play a pivotal role in the analysis of this paper. Because of FD, a node should be able to quickly detect a collision with another, if the two are covered by each other. We argue that this type of collision (we call it "covered node collision") can occur only when the nodes start transmitting at the same slot. Due to a processing overhead that exists for detecting a simultaneous transmission on the medium, we assume nodes cease their transmission after the header time which takes $\tau_V$ time slots. 

However, collision between hidden terminals is more tricky. Due to being hidden from each other, nodes cannot know of the collision by themselves. So the easy argument is that nodes know of such collision when they don't receive any BT from the AP at the end of their header. But we argue that this could potentially result in an unpredictable succession of collision by several hidden nodes without realizing that there has just been a collision in the network. For this reason we introduce a collision notification signal sent out by the AP once it detects a collision by two or more hidden nodes. Several works \cite{Sen09,Sen10} have shown that even a half duplex AP (receiver in general) can detect collision once it occurs. These works have then used out of band communication to notify the colliding nodes once they detect it. However in FD there is no need for out-of-band communication since nodes are capable of sending and receiving on the same channel simultaneously. The same scenario is  envisaged for FD nodes in \cite{Srinivasan12}. Therefore, we consider the same capability in our network setting where the AP can notify the colliding nodes immediately after detecting the collision. Once transmitting nodes are notified, they cease their transmission.

From a client's perspective, collision will also take $\tau_V$ time slots. However from the AP's perspective the length of collision on the channel is varying depending on the relative starting time of the second colliding client. If the two nodes start on the same slot, collision will take $\tau_V$ time slots. If the second node start exactly after the first node has sent it's header, then collision take $2\tau_V$ time slots. We can say that in average, hidden terminal collision will take $3\tau_V/2$ slots. % discuss how we consider this value for AP in our analysis. 
%Also talk about the small gap between header and payload due to PHY limitations, which works well for finding out if the AP has replied. 

\section{Analytical Model for Full Duplex MAC}
\label{sec:model}

\begin{figure}[!t]
\centering
\includegraphics[width=0.55\textwidth]{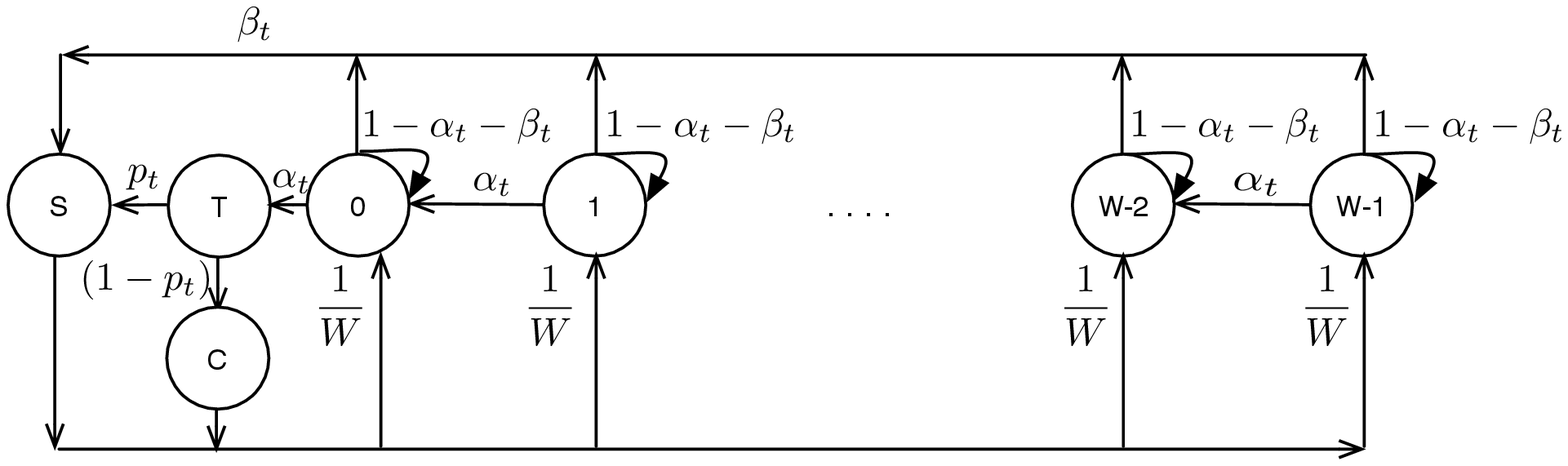}
\caption{Embedded Markov chain representing the states of head-of-line packet in full duplex enabled CSMA/CA.}
\label{fig:markov}
\end{figure}

We model the network described in Section~\ref{sec:arch} using a discrete-time Markov renewal process $M$ shown in Fig.~\ref{fig:markov}. This Markov process shows the state space of each HOL packet at the node's transmit queue. By analyzing the steady state of the Markov process, we obtain the throughput performance of nodes in the network. Our analysis is applicable for saturated conditions, i.e., every node has a packet to transmit.  

For every HOL packet, every node, either AP or client, starts from state $\mathrm{S}$. % after it has found the channel idle. 
It randomly chooses a back off counter in the range $[0,W)$ and moves to the corresponding state $0,...,W-1$ to start the countdown. From this state, the node counts down and transits to the lower state with probability $\alpha_t$, if it finds the channel idle in time slot $t$. Also let $\beta_t$ be the probability that the channel is found busy and the initial header bits are decoded to reveal that the packet is addressed to this specific sensing node, and this node's HOL packet is also for the transmitter of the header (i.e., the primary transmitter). When, this condition occurs, there is a possibility of beginning FD operation. In this case, the node immediately sends the HOL packet to the primary transmitter and then directly moves to state $\mathrm{S}$. Otherwise with probability $1-\alpha_t-\beta_t$, the node stays in the same state, and continues to sense the channel in the next time slot. 

When the node reaches state $0$, if it finds the channel idle (with probability $\alpha_t$), it attempts a transmission by sending the packet header and moving to state $\mathrm{T}$. After header transmission is completed and the transmission is successful, i.e., no collision has occurred, the node continues sending the whole packet and makes a transition to state $\mathrm{S}$. Otherwise a transition to state $\mathrm{C}$ takes place. In the latter case, the back off process is repeated to attempt a re-transmission of the collided packet. The probability of a successful transmission is given as $p_t$. 

% may be you have to mention here in what aspects this model is different from the half duplex case.

% be more elaborate, sudden jump to math is probably not a good way.
The transition probabilities in the above Markov chain can be expressed as:
\small
\begin{align}
&P[i | i+1]  =\alpha_t, P[S | i] =\beta_t , P[S | T] =p_t, \nonumber \\
&P[C | T] =1-p_t, P[i | C] =P[i | S]=\frac{1}{W} 
\end{align}
\normalsize
The Markov chain in Fig.~\ref{fig:markov} can be easily shown to be uniformly strongly ergodic if and only if the limits below exist~\cite{Dai13,Iosifescu}:
\small
\begin{equation}
\lim_{t \to \infty} p_t=p, \lim_{t \to \infty} \alpha_t=\alpha, \lim_{t \to \infty} \beta_t=\beta
\label{eq:limit}
\end{equation}
\normalsize
and therefore, has a stationary probability distribution. It is straightforward to derive the steady-state probability distribution from the following set of equations:

\small
\begin{align}
&\pi_{W-1}=\frac{1}{W(\alpha+\beta)}(\pi_{S}+\pi_C) \nonumber \\
&\pi_i=u\pi_{i+1}+\frac{1}{W(\alpha+\beta)}(\pi_{S}+\pi_C)   \nonumber \\
&0 \le i < W-1 \nonumber \\
&\pi_T=\alpha \pi_0 \nonumber \\
&\pi_{S}=p\pi_T + \beta(\sum_{j=0}^{W-1}\pi_j) \nonumber \\
&\pi_{C}=(1-p) \pi_T \nonumber \\
&\pi_{S}+\pi_C+\pi_T+\sum_{j=0}^{W-1} \pi_j = 1 
\label{eq:equset}
\end{align} 
\normalsize

where,
\begin{equation}
u=\frac{\alpha}{\beta+\alpha} \nonumber
\end{equation}

By solving (\ref{eq:equset}), the probability $\pi_S$ is obtained as:
\small
\begin{equation}
\pi_{S}=\frac{W(1-u)\beta - Xu\beta(1-p)}{W(1-u)(\beta+1) - Xu(1-\beta)} 
\end{equation}
\normalsize
where $X$ is short for:
\begin{align}
X&=(1-u^W) \\
\end{align}
\normalsize

Also, $\pi_{C}$ and $\pi_i, 0=1,...,W-1$ can be given as:
\small
\begin{align}
&\label{eq:pit} \pi_T=\frac{Xu\beta}{W(1-u)(\beta+1) - Xu(1-\beta)} \\
&\label{eq:pic} \pi_C=\frac{Xu\beta(1-p)}{W(1-u)(\beta+1) - Xu(1-\beta)} \\
&\label{eq:pii} \pi_i=\frac{(1-u)(1-u^{W-i})}{W(1-u)(\beta+1) - Xu(1-\beta)} 
\end{align}
\normalsize

The mean holding time for count down states $i=0,1,..., W-1$ can be obtained using geometric distribution as $\delta_i=\frac{1}{\alpha+\beta}$. Also assuming $\delta_T$, $\delta_S$ and $\delta_C$ as mean holding times of the states $\mathrm{T}$, $\mathrm{S}$ and $\mathrm{C}$ respectively, the limiting state probabilities for Markov process $M$ are given by:%~\cite{Dai13}

\begin{equation}
\tilde{\pi_j}=\frac{\pi_j \delta_j}{\sum_{i \in M}\pi_i \delta_i}
\label{eq:limitprob}
\end{equation}

Based on this formulation, the limiting probability of state $\mathrm{S}$, $\tilde{\pi_S}$ is the service rate or \textit{throughput} of the corresponding node's queue~\cite{Dai13}.  Now if the packet arrival rate $\lambda$ to each node's queue is more than the service rate, the node is in saturated mode. %In this paper, we are interested in the performance in this mode of operation and thus assume that all nodes are in saturated mode. 

%\begin{equation}
%\rho=\frac{\lambda}{\tilde{\pi_S}}
%\end{equation} 

The derived steady-state probability distribution is valid for all nodes, including the AP. However, the probability values are on a per-node basis, since the Markov transition probabilities for different nodes are not the same. For example, the probability of finding the channel idle $\alpha$ is different for each node since the packet arrival rate $\lambda$, and the number of covered and hidden terminals are different for each node. More specifically, client nodes are always sending packets to the AP and they only receive packets from the AP. This makes the AP's perspective of channel activity different from that of the client nodes. For the subsequent discussion, we apply the superscript per-node parameters as $\alpha^i$, $\beta^i$ and $p^i$ for node $i, i=1,...,n$ and $\alpha^{ap}$, $\beta^{ap}$ and $p^{ap}$ for the AP. We derive analytical expressions for these parameters separately for clients and the AP in the rest of this section. In Table~\ref{tab:note}, we list the commonly used notations.

\subsection{Analysis from a client's perspective}
Each node in our system has a different view of the channel when other nodes are transmitting. From an arbitrary client's perspective, say $i$ in Fig.~\ref{fig:net}, the channel can be in five different states when it is not transmitting: 
% Explain 1) which has two sub states
\begin{itemize}
\item 1) Successful FD: Client node $i$ may overhear a client $k$ within its coverage radius initiate communication to the AP, while the AP replies in FD mode (FD1 case). 
Client $i$ may also hear the AP initiating a successful FD transmission to any other client node, either covered by it ($k$) or hidden from it ($m$) (FD2 case). In all of these cases, the time taken will be $\tau_F$.  

\item 2) Successful HD:  When AP's HOL packet is not for $k$, and $k$ initiates a transmission to the AP, client $i$ observes the channel usage in the situation depicted in Fig.~\ref{fig:HD2} (HD1 case). Also, if $m$ initiates a packet transmission to the AP and a successful full duplex exchange begins between them, client $i$ (hidden from $m$) only hears one part of the communication, which is a HD transmission (HD2 case). 

\item 3) BT-ACK transmission: Client $i$ may receive only the long BT and the ACK sent by the AP, as shown in Fig.~\ref{fig:HD2}, to a hidden client (say, $m$) during and following a successful HD transmission by the client. The length of the BT-ACK is considered as $\tau_A$.

\item 4) Failed transmission: A client may sense a collision. The collision could happen in two cases: (i) when two nodes or more start transmission in the same slot, or, (ii) one or more hidden clients transmit in the vulnerable periods of a transmitting node. The length of a collision is considered to be $\tau_V$ which is the length of a packet header, as collisions are detected during time using full duplex. 
 
\item 5) Idle.
\end{itemize}

\small
\begin{table}
\caption{List of notations used in the analysis}
\begin{tabular}{| l | l |}
\hline
Symbol & Definition \\
\hline 
\hline
$\mathcal{N}$ & set of client nodes \\
$n$ & total number of client nodes or $|N|$\\
$\mathcal{N}_c^i$ & set of client nodes that are covered by node $i \in \{\mathcal{N}, \mathrm{AP}\}$ \\
$\mathcal{N}_h^i$ & set of client nodes hidden from node $i$ \\
$\omega^i(t)$ & prob. of transmission attempt by node $\mathrm{i}$ at slot $\mathrm{t}$, given that \\
& channel sensed idle at slot $\mathrm{t}-1$\\
$\omega^{ap}(t)$ & prob. of transmission attempt by the AP at slot $\mathrm{t}$, given that \\
& channel sensed idle at slot $\mathrm{t}-1$\\
$\nu^i(t)$ & prob. of transmission attempt by node $\mathrm{i}$ in the next $\tau_V$ slot \\
& until $\mathrm{t+\tau_V}$, given that channel sensed idle at slot $\mathrm{t}-1$\\
$\nu^{ap}(t)$ & prob. of hearing an AP's FD reply at $\mathrm{t}$, given that \\ & channel sensed idle at $\mathrm{t}-1$\\
$\delta_j$ & the average time spent in state $j$ known as mean holding  \\
& time of state $j$ \\
$\pi_j$ & stationary prob. of state $j$ \\
$\tilde{\pi_{\mathrm{S}}}^i$ & throughput of node $i$ equal to the limiting prob. of state $\mathrm{S}$ \\
\hline
\end{tabular}
\vspace{1mm}
\label{tab:note}
\end{table}
\normalsize

The probability of sensing the channel idle by node $i$ at each time slot $t+1$, $\alpha_{t+1}^i$ is dependent on the state of the channel at the previous time slot. Hence, we have: 
\small
\begin{align}
\alpha_{t+1}^i&=P_i[idle\  at\  t+1 | FD\  at\  t] P_i[FD\  at\  t] \nonumber \\ 
&+ P_i[idle\  at\  t+1 | HD\  at\  t] P_i[HD\  at\  t] \nonumber \\
&+ P_i[idle\  at\  t+1 | BTACK\  at\  t] P_i[BTACK\  at\  t] \nonumber \\
&+ P_i[idle\  at\  t+1 | Collision\  at\  t] P_i[Collision\  at\  t] \nonumber \\
&+ P_i[idle\  at\  t+1| idle\  at\  t] P_i[idle\  at\  t] 
\label{eq:alpha}
\end{align}
\normalsize
Considering the assumption of fixed header and packet length for all the nodes in the network, the successful FD, HD, BT-ACK and collision take respectively $\tau_F$, $\tau_H$, $\tau_A$ and $\tau_V$ slot times to finish, therefore the conditional probabilities in (\ref{eq:alpha}) would be as the following:
\small
\begin{align}
P_i[idle\  at\  t+1 | FD\  at\  t]=\frac{1}{\tau_F} \nonumber \\
P_i[idle\  at\  t+1 | HD\  at\  t]=\frac{1}{\tau_H} \nonumber \\
P_i[idle\  at\  t+1 | BTACK\  at\  t]=\frac{1}{\tau_A} \nonumber \\
P_i[idle\  at\  t+1 | Collision\  at\  t]=\frac{1}{\tau_V} 
\label{eq:cond}
\end{align}  
\normalsize

For the last conditional probability, namely $P_i[idle\  at\  t+1| idle\  at\  t]$, we use $\omega^k$, $\omega^{ap}$ and $\nu^k$ and $\nu^{ap}$ as noted in Table \ref{tab:note}:
%$\omega^k$ is the conditional probability of node $\mathrm{k}$ making transmission attempt at slot time $\mathrm{t}$, having sensed the channel at slot time $\mathrm{t}-1$. $\omega^{ap}$ is the same probability for the AP, and $\nu^k$ is the probability of at least $\tau_V$ slots away from making a transmission attempt. In other words, $\nu$ is the probability of being at a back off counter of $\tau_V$ or higher. $\nu^{ap}$ is the conditional probability that the AP responds to a transmission by a hidden node at time $\mathrm{t}$ given that the channel is sensed idle at time $\mathrm{t}-1$. 

\begin{align}
&P_i[idle\  at\  t+1| idle\  at\  t]  \nonumber \\
&=(1-\omega^{ap}(t+1)) (1-\nu^{ap}(t+1))\prod_{k \in \mathcal{N}_c^i}(1-\omega^k(t+1))
\label{eq:idle}
\end{align}

Apart from the conditional probabilities in (\ref{eq:alpha}), the probability of each channel state must be calculated. We start with the state of successful FD that is given by:
\small
\begin{align}
&P_i[FD\  at\  t] = \sum_{j=1}^{\tau_F} P_i[FD\  at\  t-j+1 | idle\  at\  t-j] P_i[idle\  at\  t-j] \nonumber \\
 \label{eq:fd}
\end{align}
\normalsize

The inner conditional probability is the summation of conditional probabilities for the two cases of FD1 and FD2 explained earlier:
\small
\begin{align}
&P_i[FD\  at\  t-j+1 | idle\  at\  t-j]=\nonumber\\
 &=P_i[FD1\  at\  t-j+1 | idle\  at\  t-j]\nonumber\\
 &+P_i[FD2\  at\  t-j+1 | idle\  at\  t-j]
 \label{eq:fdd}
\end{align}
\normalsize

For the FD1 case, the probability that client $i$ initiates a transmission to the AP accompanied by the AP's simultaneous reply in FD, is given by:
\small
\begin{align}
&P_i[FD1\  at\  t-j+1 | idle\  at\  t-j]= \nonumber \\
&\sum_{k \in \mathcal{N}_c^i} \frac{1}{n} \omega^k(t-j+1)(1-\omega^{ap}(t-j+1))\times \nonumber \\
&\prod_{l \in \mathcal{N}_c^k-i}(1-\omega^l(t-j+1)) \prod_{m \in \mathcal{N}_h^k}\nu^m(t-j+1)
\label{eq:fd1}
\end{align}
\normalsize

The terms above show the total probability for any node $k$ in the set $\mathcal{N}_c^i$ having a successful FD transmission. This requires node $k$ transmitting at time slot $t-j+1$ when the HOL packet of AP is for $k$, and the AP and all nodes in $\mathcal{N}_c^k$ are silent at that time slot and all of its hidden terminals in $\mathcal{N}_h^k$ are silent for the duration of vulnerable period $\tau_V$ after $t-j+1$. Note that we assume the AP's outgoing traffic to its clients is uniform, hence $\frac{1}{n}$ is the probability that AP's HOL packet is destined to the client node that is transmitting to it leading to FD, while probability $\frac{n-1}{n}$ is for otherwise, when the AP sends busy tone as in the Fig.~\ref{fig:HD2}.

The probability for FD2 case can also be written as: 
 \small
\begin{align}
&P_i[FD2\  at\  t-j+1 | idle\  at\  t-j]= \nonumber \\
&\omega^{ap}(t-j+1)\prod_{l \in \mathcal{N}-i} (1-\omega^l(t-j+1))
\label{eq:fd2}
\end{align}
\normalsize

 which is the total probability of all clients keeping silent, while the AP is attempting a transmission to its intended client at slot time $t-j+1$. In this case, the intended client will respond with a FD packet with probability $1$ since the client nodes are always in saturated mode, and they only send packets to the AP. 

We can formulate expressions similar to (\ref{eq:fd}) for HD and BT-ACK as well. For HD, we have the following:
 \footnotesize
 \begin{align}
&P_i[HD\  at\  t] \nonumber \\
&= \sum_{j=1}^{\tau_H} P_i[HD\  at\  t-j+1 | idle\  at\  t-j] P_i[idle\  at\  t-j]\\
&P_i[BTACK\  at\  t] \nonumber \\
&= \sum_{j=1}^{\tau_A} P_i[BTACK\  at\  t-j+1 | idle\  at\  t-j] P_i[idle\  at\  t-j]
\label{eq:hd}
\end{align}
 \normalsize

Similar to (\ref{eq:fdd}), $P_i[HD\  at\  t-j+1 | idle\  at\  t-j]$ can be written as the summation of HD1 and HD2 cases. 
\small
\begin{align}
&P_i[HD\  at\  t-j+1 | idle\  at\  t-j]=\nonumber\\
 &=P_i[HD1\  at\  t-j+1 | idle\  at\  t-j]\nonumber\\
 &+P_i[HD2\  at\  t-j+1 | idle\  at\  t-j]
 \label{eq:hdd}
\end{align}
\normalsize

For HD1 we have:
  \small
\begin{align}
&P_i[HD1\  at\  t-j+1 | idle\  at\  t-j]= \nonumber \\
&\sum_{k \in \mathcal{N}_c^i} \frac{n-1}{n}\omega^k(t-j+1)(1-\omega^{ap}(t-j+1)) \nonumber \\
&\prod_{l \in \mathcal{N}_c^k-i}(1-\omega^l(t-j+1))  \prod_{m \in \mathcal{N}_h^k}\nu^m(t-j+1)
\label{eq:hd1}
\end{align}
\normalsize
which is the total probability over any node $k$ in $\mathcal{N}_c^i$ to attempt transmission at time slot $t-j+1$, while (i) nodes in $\mathcal{N}_c^k$ and the AP keep silent at that slot, and (ii) its hidden nodes do not transmit during the next $\tau_V$ slots. This is multiplied by $\frac{n-1}{n}$, which is the probability leading up to HD1 case as explained earlier. %Note that in this case, AP's HOL packet is not for transmitting node which has the probability of $\frac{n-1}{n}$.
For HD2 case:
  \small
\begin{align}
&P_i[HD2\  at\  t-j+1 | idle\  at\  t-j]= \frac{1}{n} \nu^{ap}(t-j+1)%\nonumber \\
%&\sum_{k \in \mathcal{N}_H^i} \frac{1}{n}\omega^k(t-j-\tau_{V}+1)(1-\omega^{ap}(t-j-\tau_{V}+1))\times \nonumber \\
%&\prod_{l \in \mathcal{N}_c^{k}}(1-\omega^l(t-j-\tau_{V}+1)) \prod_{m \in \mathcal{N}_H^k-i}\prod_{s=1}^{\tau_{V}}(1-\omega^m(t-j-\tau_{V}+s)) \nonumber \\
\label{eq:hd2}
\end{align}
\normalsize

This is also the probability of hearing an AP's FD reply to a hidden node of $i$.
%Above is the sum probability for hidden nodes of $i$ within the set ${N}_H^i$ to have attempted transmission $\tau_V$ slot ago at time $t-j-\tau_{V}+1$ with a corresponding HOL packet for AP with probability $1/n$ at that time and also that its covered nodes and the AP have kept silent at that slot and all of its hidden nodes excluding the observing node $i$ to have transmitted during the vulnerable period. 

For the state of hearing the BT-ACK, similarly, we have:
 \small
\begin{align}
&P_i[BTACK\  at\  t-j+1 | idle\  at\  t-j]= \frac{n-1}{n} \nu^{ap}(t-j+1)%\nonumber \\
%&\sum_{k \in \mathcal{N}_H^i} \omega^k(t-j-\tau_{V}+1) \frac{n-1}{n}(1-\omega^{ap}(t-j+1))\times \nonumber \\
%& \prod_{l \in \mathcal{N}_c^k} (1-\omega^l(t-j-\tau_{V}+1)) \prod_{m \in \mathcal{N}_H^k-i}\prod_{s=1}^{\tau_{V}}(1-\omega^m(t-j-\tau_{V}+s)) 
\end{align}
\normalsize

which is equal to the probability of hearing a BT by the AP in response to the HD transmission of a hidden node of $i$.

For the idle state, we get $P[idle\  at\  t]=\alpha_t$. Consequently, for the state of collision sensed by node $i$, we have:

\small
\begin{align}
P_i[Collision\  at\  t] &= 1-P_i[FD\  at\  t]-P_i[HD\  at\  t] \nonumber \\ 
&-P_i[BTACK\  at\  t]-\alpha_t
\label{eq:col}
\end{align}
\normalsize

Now, $\alpha_t^i$ can be obtained by combining (\ref{eq:cond}-\ref{eq:idle}) and (\ref{eq:fd}-\ref{eq:col}).

\subsection{Analysis from the AP's perspective}
From the AP's point of view, the channel has three states when it is not transmitting:

\begin{itemize}
\item 1) Successful HD: The AP successfully receives a packet from a client $k$, when the AP's own HOL packet is destined for a node other than $k$.
\item 2) Collision: Clients that are hidden from each other could collide or two or more nodes attempt transmission exactly at the same time. 
\item 3) Idle: The channel remains unused.
\end{itemize}
% explain why we don't consider FD case in here....
A question that might arise here is that why no state for FD is considered here. The reason is, the above states are from the perspective of the AP when it is in channel contention (sensing) mode. Once AP enters an FD mode, it leaves contention mode and that is why it is not considered here. With the above  three states, we are able to derive parameters of the Markov chain $M$ for the AP.  
%Clearly, from the AP's perspective, it never observes the channel to be already present in FD mode, since client transmit to the AP only, and not to each other. 

Now, similar to (\ref{eq:alpha}), we have:
\small
\begin{align}
\alpha_{t+1}^{ap}&=P_{ap}[idle\  at\  t+1 | HD\  at\  t] P_{ap}[HD\  at\  t] \nonumber \\
&+ P_{ap}[idle\  at\  t+1 | Collision\  at\  t] P_{ap}[Collision\  at\  t] \nonumber \\
&+ P_{ap}[idle\  at\  t+1| idle\  at\  t] P_{ap}[idle\  at\  t] 
\label{eq:alphaap}
\end{align}
\normalsize

where $P_{ap}[idle\  at\  t+1 | HD\  at\  t]$ and $P_{ap}[idle\  at\  t+1 | Collision\  at\  t]$ are the same as given in (\ref{eq:cond}). 

Also, $P_{ap}[idle\  at\  t+1| idle\  at\  t] $ is exactly $p_t^{ap}$, as defined in Section \ref{sec:model} and is obtained as:
\begin{align}
&P_{ap}[idle\  at\  t+1| idle\  at\  t] = p_t^{ap} \nonumber \\
&=\prod_{k=1}^{n}(1-\omega^k(t+1))
\label{eq:ptap}
\end{align}

This is the probability of a successful AP transmission in time slot $t$, if a transmission attempt is made, and given the channel is idle at slot time $t-1$.

The probability that the AP senses the channel in HD state is given by:

\small
\begin{equation}
P_{ap}[HD\  at\  t] = \sum_{j=1}^{\tau_H} P^{ap}[HD\  at\  t-j+1 | idle\  at\  t-j] P^{ap}[idle\  at\  t-j]
%\label{eq:hd}
\end{equation}
\normalsize

where,
 \small
\begin{align}
&P_{ap}[HD\  at\  t-j+1 | idle\  at\  t-j]= \nonumber \\
&\sum_{k=1}^{n}  \frac{n-1}{n}\omega^k(t-j+1) \times \nonumber \\
& \prod_{l \in \mathcal{N}_c^k}(1-\omega^l(t-j+1))  \prod_{m \in \mathcal{N}_h^k}\nu^m(t-j+1)
\end{align}
\normalsize
The latter is the total probability of any client $k$ transmitting to AP at slot $t-j+1$, when AP's HOL packet's destination is not for $k$, and the following conditions hold: (i) AP and nodes in $\mathcal{N}_c^k$ keep silent in that slot, and (ii) the nodes in $\mathcal{N}_h^k$ keep silent for at least the next $\tau_V$ slots.
  
For the state of collision, similar to (\ref{eq:col}) we get: 
\small
\begin{equation}
P_{ap}[Collision\  at\  t] = 1-P_{ap}[HD\  at\  t]-\alpha^{ap}_t
\label{eq:colap}
\end{equation}
\normalsize
Now, $\alpha_t^{ap}$ can be obtained by substituting (\ref{eq:ptap}-\ref{eq:colap}) in (\ref{eq:alphaap}).
\subsection{Steady-state probabilities}
%It is relatively easy to show from (\ref{eq:alpha}-\ref{eq:colap}) that 
Due to symmetry, if the following conditions hold, then as $t \to \infty$, in the steady state we have $\omega^i=\omega, \nu^i=\nu, i=1...n$. 

\begin{align}
&  |\mathcal{N}_c^i|=n_c, |\mathcal{N}_h^i|=n_h,
&i=1,...,n
\label{eq:conditions}
\end{align}

These values, as well as $\omega^{ap}$, can be written in terms of the steady state probabilities of the states of the Markov chain in Section~\ref{sec:model}. Based on the definition in Table~\ref{tab:note}, $\nu$ is the given probability that a node does not attempt a transmission in the next $\tau_V$ slots. Also $\omega$ and $\omega^{ap}$ are the respective probabilities for the node and AP for attempting a transmission in an arbitrary slot, which is equivalent to entering state $T$.
\begin{equation}
\begin{cases}
& \nu=\sum_{j=\tau_V}^{W-1}\pi_j \\
& \omega^{ap} = {\pi_T^{ap}} \\
& \omega={\pi_T}
\label{eq:wv}
\end{cases}
\end{equation}
Also according to the definitions in Table~\ref{tab:note}, $\nu^{ap}$ is the probability that a node only hears an FD reply by the AP. This event is triggered when a hidden client sent a successful packet to the AP. In the steady state, this probability can be expressed as the event that any of the hidden terminals of an arbitrary client attempts a transmission that subsequently succeeds:
\begin{equation}
\nu^{ap}=n_H\omega(1-\omega^{ap})(1-\omega)^{n_c}\nu^{n_h}
\end{equation}

Once the conditions in (\ref{eq:conditions}) are met, the terms of (\ref{eq:alpha}) in the steady state can be simplified as the following (\ref{eq:cond}-\ref{eq:col}):
\small
\begin{align}
&P[idle\  at\  t+1 | FD\  at\  t] P[FD\  at\  t]_{t \to \infty} \nonumber \\
&=\frac{1}{\tau_F}\tau_F[ \frac{n_c}{n}\omega(1-\omega^{ap})(1-\omega)^{n_c-1}\nu^{n_h} + \omega^{ap}(1-\omega)^{n-1} ]\alpha \nonumber \\
& = Y_1\alpha \label{eq:y1} \\
%\end{align}
\nonumber \\
%\begin{align}
&P[idle\  at\  t+1 | HD\  at\  t] P[HD\  at\  t]_{t \to \infty} \nonumber \\
&=\frac{1}{\tau_H}\tau_H[ n_c\frac{n-1}{n}\omega(1-\omega^{ap})(1-\omega)^{n_c-1}\nu^{n_h}  \nonumber \\
&+\frac{n_h}{n}\omega(1-\omega^{ap})(1-\omega)^{n_c}\nu^{n_h-1} ] \alpha=Y_2\alpha \\
%\end{align}
\nonumber \\
%\begin{align}
&P[idle\  at\  t+1 | BTACK\  at\  t] P[BTACK\  at\  t]_{t \to \infty} \nonumber \\
&=\frac{1}{\tau_A}\tau_A[n_h\frac{n-1}{n}\omega(1-\omega^{ap})(1-\omega)^{n_c}\nu^{n_h-1} ]\alpha \nonumber \\
&=Y_3\alpha \\
%&\alpha=\frac{1}{\tau_F}\sum_{j=1}^{\tau_F}
%\end{align}
\nonumber \\
%\begin{align}
&P[idle\  at\  t+1 | Collision\  at\  t] P[Collision\  at\  t]_{t \to \infty} \nonumber \\
&=\frac{1}{\tau_C}[1-\tau_FY_1\alpha-\tau_HY_2\alpha-\tau_AY_3\alpha-\alpha] \\
%\end{align}
\nonumber \\
%\begin{align}
&P[idle\  at\  t+1 | idle\  at\  t] P[idle\  at\  t]_{t \to \infty} \nonumber \\
&=(1-\omega)^{n_c}(1-\omega^{ap})[1-n_H\omega(1-\omega^{ap})(1-\omega)^{n_c}\nu^{n_h-1}]\alpha \nonumber \\
&= Y_4\alpha \label{eq:y5}
\end{align}
\normalsize

By replacing (\ref{eq:y1}-\ref{eq:y5}) in (\ref{eq:alpha}), $\alpha$ can be obtained as:
\small
\begin{equation}
\alpha=\frac{1}{1+(\tau_F-\tau_C)Y_1+(\tau_H-\tau_C)Y_2+(\tau_A-\tau_C)Y_3+\tau_C(1-Y_4)}
\label{eq:alphatot}
\end{equation}
\normalsize

For the AP in the steady state, we get from (\ref{eq:alphaap}-\ref{eq:colap}):
\small
\begin{align}
&P^{ap}[idle\  at\  t+1 | HD\  at\  t] P^{ap}[HD\  at\  t]_{t \to \infty}\nonumber\\
&=(n-1)\omega(1-\omega)^{n_c}\nu^{n_h}\alpha^{ap} \nonumber\\
&=Z_1\alpha^{ap} \label{eq:z1} \\
%\end{align}
\nonumber \\
%\begin{align}
&P[idle\  at\  t+1 | Collision\  at\  t] P[Collision\  at\  t]_{t \to \infty} \nonumber \\
&=\frac{1}{\tau_C}[1-\tau_HZ_1\alpha^{ap}-\alpha^{ap}] \\
%\end{align}
\nonumber \\
%\begin{align}
&P[idle\  at\  t+1 | idle\  at\  t] P[idle\  at\  t]_{t \to \infty}\nonumber \\
&=p^{ap}\alpha^{ap} =(1-\omega)^{n}\alpha^{ap}  \label{eq:z3}
\end{align}
\normalsize

By replacing (\ref{eq:z1}-\ref{eq:z3}) in (\ref{eq:alphaap}), $\alpha^{ap}$ can be obtained as:
\small
\begin{equation}
\alpha^{ap}=\frac{1}{1+(\tau_H-\tau_C)Z_1+\tau_C(1-p^{ap})}
\label{eq:alphaaptot}
\end{equation}
\normalsize

Furthermore, the probability of successful transmission for a client ($p$) can be written as the product of the probabilities that (i) the AP and the other covered client nodes do not transmit, and (ii) the AP does not have to respond in FD mode to any transmission initiated by any of the remaining hidden nodes: 
\begin{equation}
p=(1-\omega^{ap})(1-\omega)^{n_c}\nu^{n_h}
\end{equation}

Also, $\beta$ and $\beta^{ap}$ in the steady state are given by:
\begin{align}
&\label{eq:beta} \beta=\frac{1}{n}\omega^{ap}(1-\omega)^{n-1} \\
&\label{eq:betaap} \beta^{ap}=n\frac{1}{n}\omega(1-\omega)^{n_c}\nu^{n_h}
\end{align}
For $\beta$, i.e., the probability for a client to go to FD mode during back off, we need the AP to transmit to that node. This probability is given by the product of $\frac{1}{n}\omega^{ap}$ with the probability that no other node transmits (excluding collision cases).

To calculate $\beta^{ap}$, i.e., the probability for the AP to respond in FD to a packet during back-off, (i) a client must  transmit, and (ii) the AP's HOL packet must also be for the same client (prob. $\frac{1}{n}$). To exclude collision cases, this is multiplied with the probability that no hidden node has already initiated a header transmission, and no covered node is attempting transmission in the same time slot. Overall, this is true for {\em any} of the $n$ client nodes. 

Now, replacing (\ref{eq:alphatot}), (\ref{eq:alphaaptot}-\ref{eq:betaap}) in (\ref{eq:pit}-\ref{eq:pii}), a system of non-linear equations based on (\ref{eq:wv}) is obtained. These equations must be solved for $\omega$ and $\omega^{ap}$ and $\nu$, which subsequently leads to the calculation of  $\alpha$, $\beta$, $p$, $\alpha^{ap}$, $\beta^{ap}$ and $p^{ap}$.

\subsection{Throughput Analysis}
As mentioned in section~\ref{sec:model}, the throughput is given by the limiting state probability of state $\mathrm{S}$, $\tilde{\pi_S}$. 

To calculate the average throughput of clients and the AP, $\tilde{\pi_S}$ for each must be calculated using (\ref{eq:limitprob}). The key parameters for this are the holding times of successful transmission and collision which are given by % needs more convincing
\begin{align}
&\delta_T=H, \delta_S^{ap}=\tau_F-H \\
&\delta_S=\frac{1}{n}(\tau_F-H) + (1-\frac{1}{n})(\tau_H-H), \\
&\delta_C=\frac{n_h}{n+1}\frac{\tau_{V}}{2}+\frac{n_c+1}{n+1}\sigma,\\  &\delta_C^{ap}=\sigma
\end{align}

where $\tau_F$, $\tau_H$ and $\tau_{V}$ are given by

\begin{align}
\begin{cases}
\tau_F=2H+L_p+SIFS+ACK \\
\tau_H=H+L_p+SIFS+ACK \\
\tau_{V}=H
\end{cases}
\end{align}

Here, $H$ and $L_p$ are the times to send the header and payload, respectively. Every node spends time to first send the header in state $T$ every time it attempts a transmission. A subsequent successful transmission for the AP $\delta_S^{ap}$ will always involve the remaining portion of an FD packet. However for a client this time could be that required to transmit either (i) an FD packet without the header (header is considered in state $T$) in $\frac{1}{n}$ of the cases, or (ii) an HD packet without the header in the rest of the cases. Also, for collision state $C$, the AP only spends a slot time as it knows about collision after its header is transmitted. This is also true for clients when they collide with their covered nodes. However, when a collision with hidden nodes occurs, the colliding nodes only know when they receive the notification from the AP. On average, this step takes $\frac{\tau_{V}}{2}$ time. An average of the two cases is considered for $\delta_C$ above.     

For the calculation of $\alpha$ and $\alpha^{ap}$ in (\ref{eq:alphatot}) and (\ref{eq:alphaaptot}), $\tau_C$ is used in both cases, which is the time that channel is sensed in the state of collision from a node's perspective. For both AP and clients, this state can be a combination of sub-states when collision occurs between two covered nodes ($H$) or two hidden nodes ($\frac{3H}{2}$ in average). The following expressions for $\tau_C$ are derived from simple probabilistic manipulations.

\begin{align}
&\tau_C^{ap}=\frac{n_cn}{n(n-1)}H + \frac{n_hn}{n(n-1)}\frac{3H}{2}, \label{eq:taua1} \\
&\tau_C=\frac{n_c^2+2n_cn_h+2*n_h}{n(n-1)}H + \frac{n_c^2-2n_c}{n(n-1)}\frac{3H}{2}
\label{eq:taua2}
\end{align}

In (\ref{eq:taua1}), the first term accounts for the case where the AP sees a collision occurring between a node and one of its covered nodes. Also the second term indicates the case where AP witnesses a node and ones of its hidden terminals collide.

From a client's perspective, there are three different collision cases: (i) between two covered node that at least one of them is covered by the client (collision time $H$), (ii) between two hidden nodes that at least one of them is covered by the client (collision time $\frac{3H}{2}$), (iii) collision between nodes that are hidden to the client (collision time 0). These three cases are considered in \ref{eq:taua2}.

\vspace{-5mm}
\section{Model Validation}
\label{sec:validation}

In this section, we first present extensive simulation results to validate our model for the FD MAC protocol in Section~\ref{sec:model}. For various network settings, we compare the model predictions with the simulation results and show that they are in close agreement. We then analyze and discuss quantitatively how much benefit FD ushers in over the more common HD transmission scenarios.

\begin{table}[!t]
\renewcommand{\arraystretch}{1.2}
\caption{Simulation Parameters}
\label{tab:dcfparam}
\centering
\begin{tabular}{l||l}
\hline
\bfseries Simulation Parameter & \bfseries Default Value\\
\hline\hline
%Preamble length & $144$ bits \\
%PLC header length & $48$ bits \\
MAC header &  $28 $ bytes  \\
PHY header & $24$ bytes  \\
ACK & $38$ bytes \\
Payload size & $1000$ bytes \\
Slot time &  $20$ $\mu$s  \\
%DIFS & $50$ $\mu$s  \\
SIFS & $10$ $\mu$s  \\
Channel bit rate & $1\ \mathrm{Mbps}$ \\
%ACK timeout & 500 $\mu$s   \\
%CW$_{min}$ & $32$ \\
%CW$_{max}$ &  $1024$ \\
%Retry limit &  $7$ \\
\hline
\end{tabular}
\end{table}

\subsection{Simulation Setup}

We have implemented the FD MAC protocol in the ns-2 simulator~\cite{ns2}. Each client is driven into saturation by generating CBR packets at 1Mbps. The MAC layer rate is set at 10 Mbps and the packets generated are each 1000 bytes, which after inclusion of the MAC PLCP header and the PHY preamble, have a total size of 1052 bytes. We consider the fact that the PHY preamble is always transmitted at 1Mbps when calculating the total duration of a packet transmission. We also assume symmetric FD link in terms of data rate and ignore channel errors and the capture effect.

The clients are deployed in the transmission range of an access point, each having the sensing and transmission radius of $150\ \mathrm{m}$. Additional simulation parameters used in the simulation are shown in Table~\ref{tab:dcfparam}.

All the simulation results presented here have $95\%$ confidence interval within $1\%$ relative error. The simulations in ns-2 are conducted for both ring and random network topologies as the most relevant scenarios for full duplex networks.

We evaluate the ring topology in order to merely study the hidden terminal effect on the performance of full duplex. In this topology, the average throughput of all clients are the same due to symmetry of network. In addition, the number of hidden stations can be easily controlled by varying the ring radius.
On the other hand, for more realistic cases of random topologies, we study the performance of full duplex in networks with non equal client-to-AP distances and hence, different number of hidden stations for each client across the network. 

In both cases, we vary the (i) number of hidden terminals, (ii) the number of clients, (iii) and the size of back-off window. In particular, we analyze the effect of such variations on saturation throughput of the network. We normalize the saturation throughput for each experiment as the nominal performance metric. Normalization is done over the MAC data rate which is considered to be 10Mbps in our experiments.  

%\begin{figure}[!t]
%\centering
%\includegraphics[width=0.45\textwidth]{Figs/topology2}
%\caption{Ring topology used for simulation with $n=8, n_c=2, n_H=5$, $\mathcal{N}_c^{n_1}=\{n_2,n_8\}$, and $\mathcal{N}_H^{n_1}=\{n_3,n_4,n_5,n_6,n_7\}$.}
%\label{fig:topology}
%\end{figure} 

\subsection{Simulation Results}

\subsubsection{Ring Topology}
\label{sec:ring}
At the first step, we have simulated a network of full-duplex nodes with the ring topology to meet the network conditions in~(\ref{eq:conditions}). In this topology, clients are placed uniformly at equal distance from the AP in a circle. Therefore, the number of hidden terminals for all clients can be fixed by properly adjusting the radius of the circle ($R$) while keeping the transmission range constant. %With simple geometric analysis of such topology, it can be easily derived that the number of hidden terminals cannot be more than $n-1-2\floor{\frac{n}{6}}$, $n$ being the number of clients. 

\begin{figure}[!t]
\centering
\includegraphics[width=0.5\textwidth]{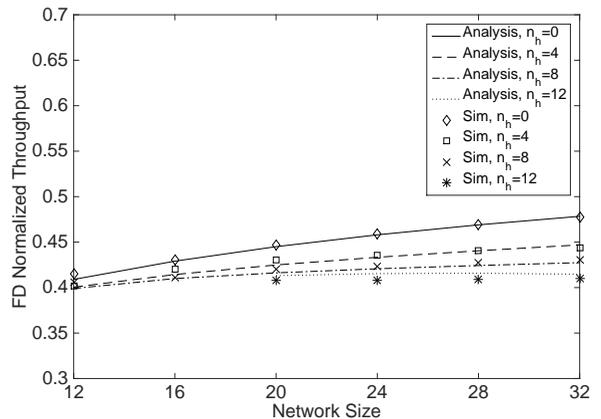}
\caption{The effect of number of clients and hidden terminals on system throughput for the ring topology.}
\label{fig:thrnnh1}
\end{figure}

Fig.~\ref{fig:thrnnh1} shows the normalized system throughput for such FD network as the number of clients increases. Throughput is shown for $n_h=0,4,8,12,16$ and $W=1024$ is considered. When nodes  are close to the AP, namely hidden terminals are low, throughput is slowly going up versus number of clients. When clients are located at the edge of AP's range, namely more hidden terminals for each client, the number of clients start to have an adverse effect on the throughput. This is visible in the case of $n_h=12$ where within the shown range, throughput is almost steady. 
%This breakage point occurs when the effect of collision probability $p_c=1-p$ outweighs probability of transmission $\omega$ in the network. Fig.\ref{fig:pnnh} show this effect for these two probabilities for networks of different sizes. 

%\begin{figure}[!t]
%	\centering
%	\includegraphics[width=0.5\textwidth]{Figs/3}
%	\caption{The effect of number of clients and hidden terminals on probability of collision.}
%	\label{fig:pnnh}
%\end{figure} 

\begin{figure}[!t]
\centering
\includegraphics[width=0.5\textwidth]{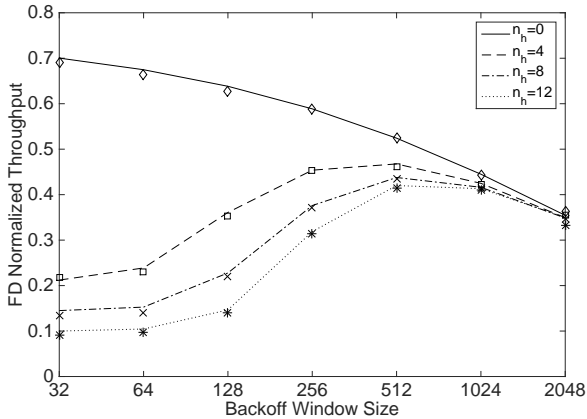}
\caption{The effect of contention window size on system saturation throughput for the ring topology.}
\label{fig:thrWnh}
\end{figure}

%Another phenomenon observed from Figs.~\ref{fig:thrnnh1} and \ref{fig:thrnnh2} is the windening gap of network throughput when no hidden terminal is present as opposed to when hidden terminals are introduced. This means that as backoff window length tightens, the effect of hidden terminal is more visible on the network performance. 
The effect of backoff window size on system throughput is shown in Fig.~\ref{fig:thrWnh} where $n=20$ is assumed. The interesting observation from this figure is a relatively large effect of hidden terminals on throughput in lower backoff sizes. But as the backoff window is enlarged this effect is diminishing to very small amount in large backoff window as in $W=2048$. This effect is attributed to the decrease in the probability of collision due to hidden terminals given a fixed network size.  
%Interestingly at $W=2048$ the throughput curves converge. This means when the window is too large, the effect of hidden terminals is very little. Continuing this trend at larger windows, the curves will cross each other and their order will reverse, although they stay very close to each other. Although due to very small difference this effect can not be seen here. The reason for this cross could be attributed to the fact that when ...% give a mathematical hint as to why this is happening.

Another observation is that, in all cases of hidden terminal presence, i.e. $n_h=4,8,12$, the network experiences a peak in throughput at $W=512$. This is due to an inherent trade off between probability of transmission attempt, channel idle time and probability of collision. In small backoff windows, probability of transmission attempt is low as channel is sensed busy most of the time and when it is not, a collision due to hidden terminals is very likely. As the backoff window is enlarged, idle channel is more common and hidden terminal collision is less likely. But after a certain point nodes start to lose idle channel opportunities by spending too much time for backoff. For $n_h=0$ the probability of collision is much smaller than when $n_h>0$, hence no peak is seen in this figure. However in smaller window sizes such peak appears for $n_h=0$ as well especially when network size is large. We do not investigate this here as it is widely studied in the other works~\cite{bianchi,Dai13}.

\subsubsection{Random Topology}
\label{sec:random}
We have extended our ns-2 simulation to include more general topologies with nodes randomly deployed around the AP. The same transmission parameters are used as in Table~\ref{tab:dcfparam}. In this deployment we use uniform distribution to place clients within the transmission range of the AP. As the density of the nodes increases, the average number of hidden terminals for each client also increases. For each random topology we have recorded the exact number of hidden terminals for each client. To be able to predict the throughput results for such general topology from the analysis given in Sec.~\ref{sec:model} we have used a similar method as in~\cite{Jang12}. Jang et al. has used the "back-of-the-envelope" approximation technique from ~\cite{Liew10} to apply their analysis of a specific topology in a network with hidden terminals to more general topologies, a situation that is applicable to our analysis as well. Such approximation has been shown to yield reasonable accuracy and works as the following:
\begin{itemize}
\item For each network node $i$ in a random topology, the number of hidden terminals ($n_h^i$) and covered nodes ($n_c^i$) are used to derive a system througput value from the analysis available for the simplified topology.
\item The system throughput of the random topology is approximated as the average of system throughput dervied from each pair of $(n_h^i, n_c^i)$ above. 
\end{itemize}

\begin{figure}[!t]
	\centering
	\includegraphics[width=0.5\textwidth]{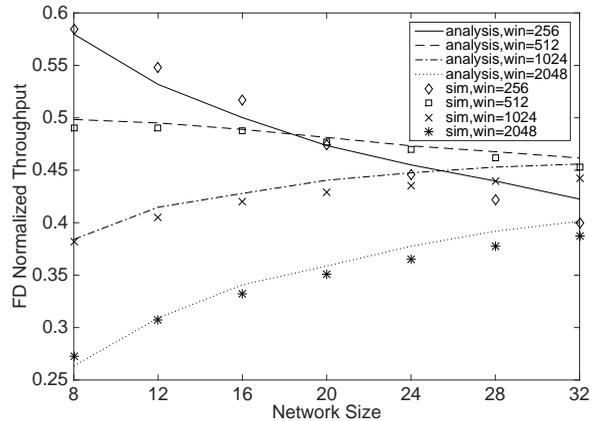}
	\caption{System throughput versus network size in random topology}
	\label{fig:random1}
\end{figure}

\begin{figure}[!t]
	\centering
	\includegraphics[width=0.5\textwidth]{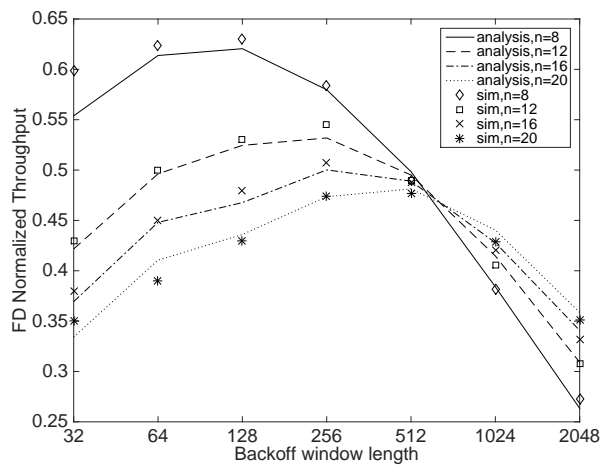}
	\caption{System throughput versus contention window in random topology}
	\label{fig:random2}
\end{figure}

%Based on this we have used the number of hidden terminal of each node in the random topology and run it through our analysis to get per-node throughput and sum over all to calculate the aggregate throughput. 
Using this method, we obtained a good approximation of the system throughput  which is shown in Figs.~\ref{fig:random1} and~\ref{fig:random2}. 
In Fig.~\ref{fig:random1} throughput versus network size is plotted for different contention window sizes. The trend in each of the curves is different depending on the contention window size. For lower value of $W=256$, system throughput is decreasing versus the network size (together with hidden terminals since it's average is also going up). For $W=512$, the throughput remains almost steady whereas for higher values of W, the trend reverses to increasing. This effect is better shown in Fig.~\ref{fig:random2} where contention window size is varied to obtain throughput. We see the rise of system throughput for various network sizes to a peak at successive windows sizes and a subsequent decline. Furthermore, we see a crossing point at $W=512$ after which larger networks outperform smaller networks. The peak phenomenon is very similar to Fig.~\ref{fig:thrWnh} where we show it is caused by hidden terminals. In that figure, we have the peak $n_h=4,8,12$ at $W=512$. Here, by going through the trace of our random topologies, we have obtained average number of hidden terminals in the network for each case of $n=8,12,16,20$ to be $n_h=0.3,1.5,2.4,3.8$ respectively.

\normalsize
\subsection{Full duplex Gain Evaluation}
To understand to what extent FD capability will improve network performance in contrast to HD, we performed a separate analysis and also conducted simulations for HD networks as well. We specifically considered HD with RTS/CTS mechanism as it is comparable to FD for its capability to mitigate hidden terminal problem. We did not consider HD basic access mechanism since its performance versus RTS/CTS is widely studied in the literature~\cite{bianchi,Dai13}. However, our analytical model can be easily extended to such scenario as well. To have similar analysis of RTS/CTS to our FD analysis, we have simplified the Markov chain in Fig.~\ref{fig:markov} by eliminating the transitions for FD transmission. We subsequently calculated the probability $\alpha$ for clients and the AP based on which we eventually derived HD throughput. We do not present this analysis here due to the lack of space, and since it can be similarly derived as in Sec.~\ref{sec:model}. We have also changed the baseline ns-2 implementation for 802.11 with RTS/CTS and adjusted it to our Markov model. Through simulations we have confirmed a match with our analysis similar to FD in sections~\ref{sec:ring} and~\ref{sec:random}. This also shows the applicability of our model to half-duplex CSMA/CA-based networks with hidden terminals. % also show if matches with simulation, also mention the time values (collision, successful tx, etc.) you used in comparison with the FD.
We analytically calculated the FD gain as the ratio of FD over HD throughput in various random topologies, obtained using the procedure in section \ref{sec:random}, to characterize its behavior in various network sizes and contention window lengths. Additionally, we have analyzed the effect of hidden terminals on FD gain in the ring topology. 

\subsubsection{Network size}
\begin{figure}[!t]
	\centering
	\includegraphics[width=0.5\textwidth]{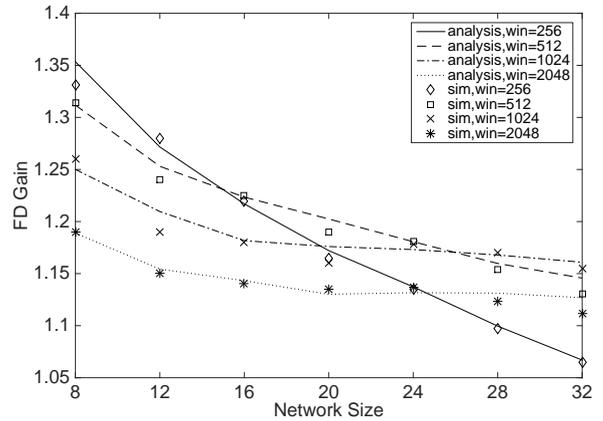}
	\caption{FD gain versus network size}
	\label{fig:gnhW}
\end{figure}

%In all cases we see improvement over half duplex with RTS/CTS and this improvement increases as the number of hidden terminal increases.
In this section, we analyze the effect of network size on the performance of FD. Fig.~\ref{fig:gnhW} shows FD gain versus network size in random topologies. The number of nodes is varied from 8 to 32. The results are shown for various contention window sizes, i.e. 256, 512, 1024 and 2048 to show their effect on the FD gain as well. 

\textit{Discussion:} As we see for all cases, gain is dropping as network becomes larger. This decreasing trend for enlarged network sizes is mainly due to HOL blocking. Remember that in FD MAC, the AP replies in FD mode to a client's transmission if its HOL packet is in fact destined for that specific client. Also for smaller contention window sizes, gain is dropping at a faster rate than larger window sizes. This is due to the dependency of a FD transmission to collisions. In other words, referring to (\ref{eq:beta}), as network becomes larger, the probability of collision increases as well, not only due to more nodes but also due to more occurrence of hidden terminal problem in random topologies. This increase in probability of collision is more severe in smaller window sizes, hence decreasing the chance of FD. Eventually, as the network becomes very large, probability of FD due to a client's transmission attempt saturates at a positive gain due to transmissions by AP which are always replied in FD by the receiving client. Note that, transmissions by the AP are much more likely to be successful because it has no hidden terminals.

\subsubsection{Contention window size}
\begin{figure}[!t]
	\centering
	\includegraphics[width=0.5\textwidth]{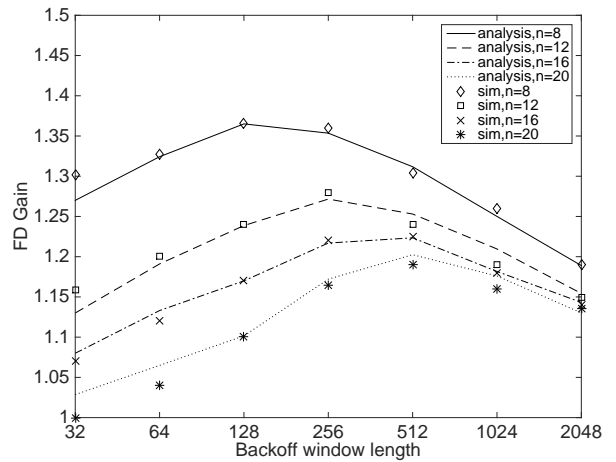}
	\caption{FD gain versus backoff window length}
	\label{fig:gWnh}
\end{figure}

The effect of contention window size on the FD gain is better shown in Fig.~\ref{fig:gWnh}. In this figure, we have plotted FD gain versus contention window size from 32 to 2048. The results are shown for four different network sizes of 8, 12, 16 and 20 nodes. 

\textit{Discussion:} In Fig.~\ref{fig:gWnh}, we see the maximum FD gain for each network size occurs at a certain contention window size. As we noted earlier, for a certain network size, a smaller window causes too many collisions and a larger window causes less transmission attempts. Therefore for each network size there is an optimum window size. Moreover, we observe gain peaks successively occur at larger window sizes as network becomes larger. %This leads us to the conclusion that basically for each network size there is a specific window in which an FD capable network benefits the most. 
In other words, as we increase the size of the network, the optimal window size increases as well. Comparing this with Fig. \ref{fig:random2}, we also see a correspondence between FD maximum throughput and gain as well. 

Another observation from Fig.~\ref{fig:gWnh} is the higher contrast of gain difference at the various network sizes in smaller windows. This shows that as transmission attempts become less likely with longer backoffs, the FD gain becomes less dependent on the network size. For sparser networks, e.g. $n = 8$, small window translates to higher transmission probability. On the other hand, for denser networks such as $n = 20$, higher transmission probability is neutralized by less FD occurrences due to HOL blocking.

We study the effect of hidden terminal on FD gain in the next section. 

\subsubsection{Hidden terminals}
\begin{figure}[!t]
\centering
\includegraphics[width=0.5\textwidth]{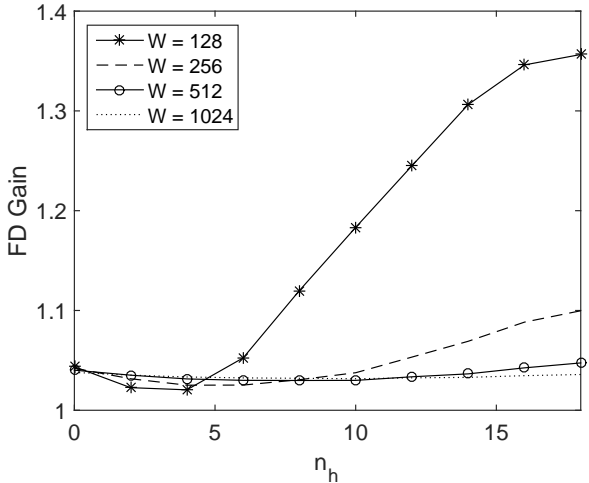}
\caption{FD gain versus number of hidden terminals in ring topology}
\label{fig:gNnh}
\end{figure}

In random topologies, the pure effect of hidden terminal cannot be clearly observed, as each node might have a different number of hidden terminals. For this purpose, we use the ring topology as in section \ref{sec:ring} where the number of hidden terminals is a controlled variable. Fig.~\ref{fig:gNnh} shows FD gain versus the number of hidden terminals varying from 0 to 18. The results are shown for a network of 30 nodes and windows of length 128, 256, 512 and 1024. 

\textit{Discussion:} The number of hidden terminals in the network has a rather peculiar effect on the gain. From the figure, we observe that the gain slightly drops up to a certain number of hidden terminals and then goes up. For larger windows, the gain is rather steady and this effect is not as visible. But this increase is more significant for lower window sizes. For $W = 128$, we see a much higher gain for large hidden terminals. %To investigate this more, we resort to a an approximate measure of gain.  

Having a closed-form expression of the FD gain could help to analyze the foregoing effect of hidden terminals on gain. But such an expression could be very complex, given two separate Markovian models of FD and HD, each leading to different parameters ($\omega$, $\omega^{ap}$, $\nu$) used in throughput derivation. However, we propose an approximate measure, that produces the same behavior as gain. Using only the solution of the FD Markovian model, we can account for the probabilities of FD successful transmissions with respect to all successful transmissions in the network (see Fig.~\ref{fig:FD}) and find the ratio in which FD occurs versus all FD and HD transmissions. In short, we have

\begin{equation}
\tilde{G} = 1 + \frac{\omega p + \omega^{ap}p^{ap}}{n\omega p + \omega^{ap}p^{ap}}
\label{eq:gain_app}
\end{equation}

As discussed in section~\ref{sec:model}, $\omega^{ap}p^{ap}$ is the probability of successful transmission by the AP which always lead to an FD reply by the receiver client. Also, $\omega p$ is the probability of a successful transmission of an arbitrary client to the AP followed by AP's FD reply with probability $\frac{1}{n}\omega p$ times $n$, the total number of clients. In the denominator of the above metric, we have $n\omega p$ which is the probability of successful transmission by all clients summed by successful probability of AP transmission $\omega^{ap}p^{ap}$. Therefore,  $\tilde{G}$ shows an estimate of FD throughput gain over HD in the same network. In the following, we use this estimate measure to explain the behavior of FD gain under various hidden terminal scenarios as seen in Fig.\ref{fig:gNnh}.

\begin{figure*}%
	\centering
	\subfigure[$\omega$]{{\includegraphics[width=0.38\textwidth]{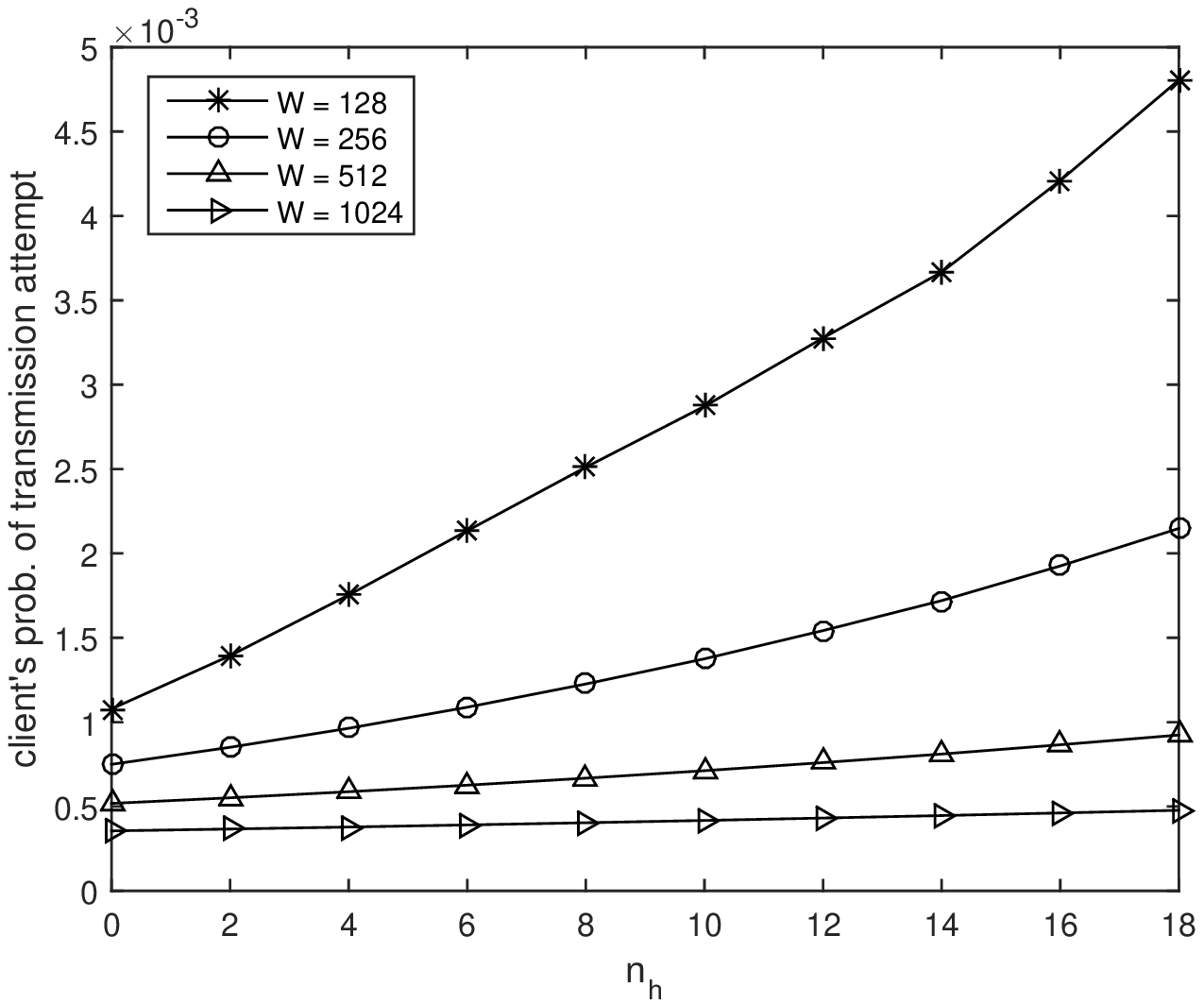} }}%
	\qquad
	\subfigure[$p$]{{\includegraphics[width=0.38\textwidth]{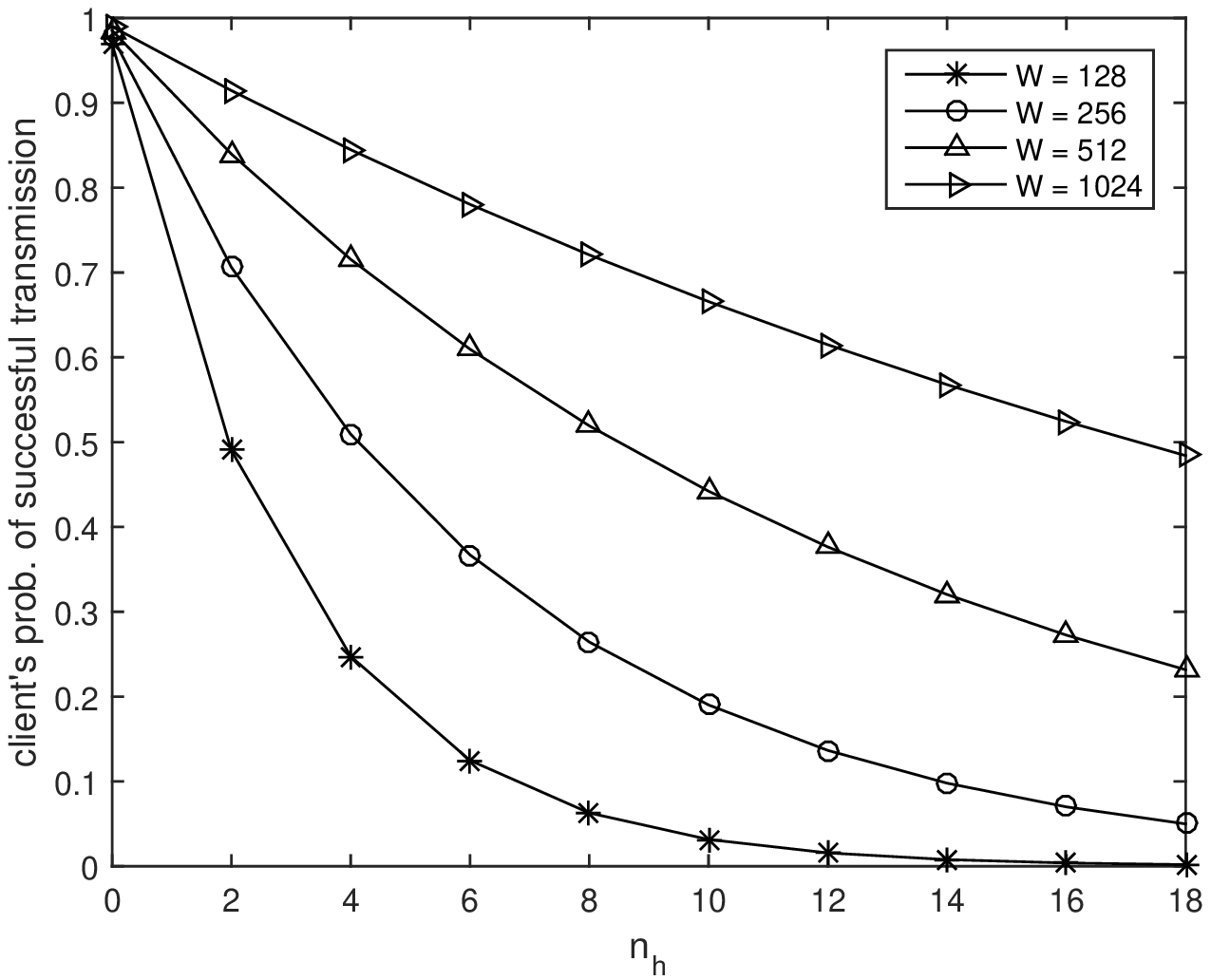} }}%
	\centering
	\caption{The effect of hidden terminals on $\omega$ and $p$}%
	\label{fig:example1}%
\end{figure*}

\begin{figure*}%
	\centering
	\subfigure[$\omega^{ap}$]{{\includegraphics[width=0.38\textwidth]{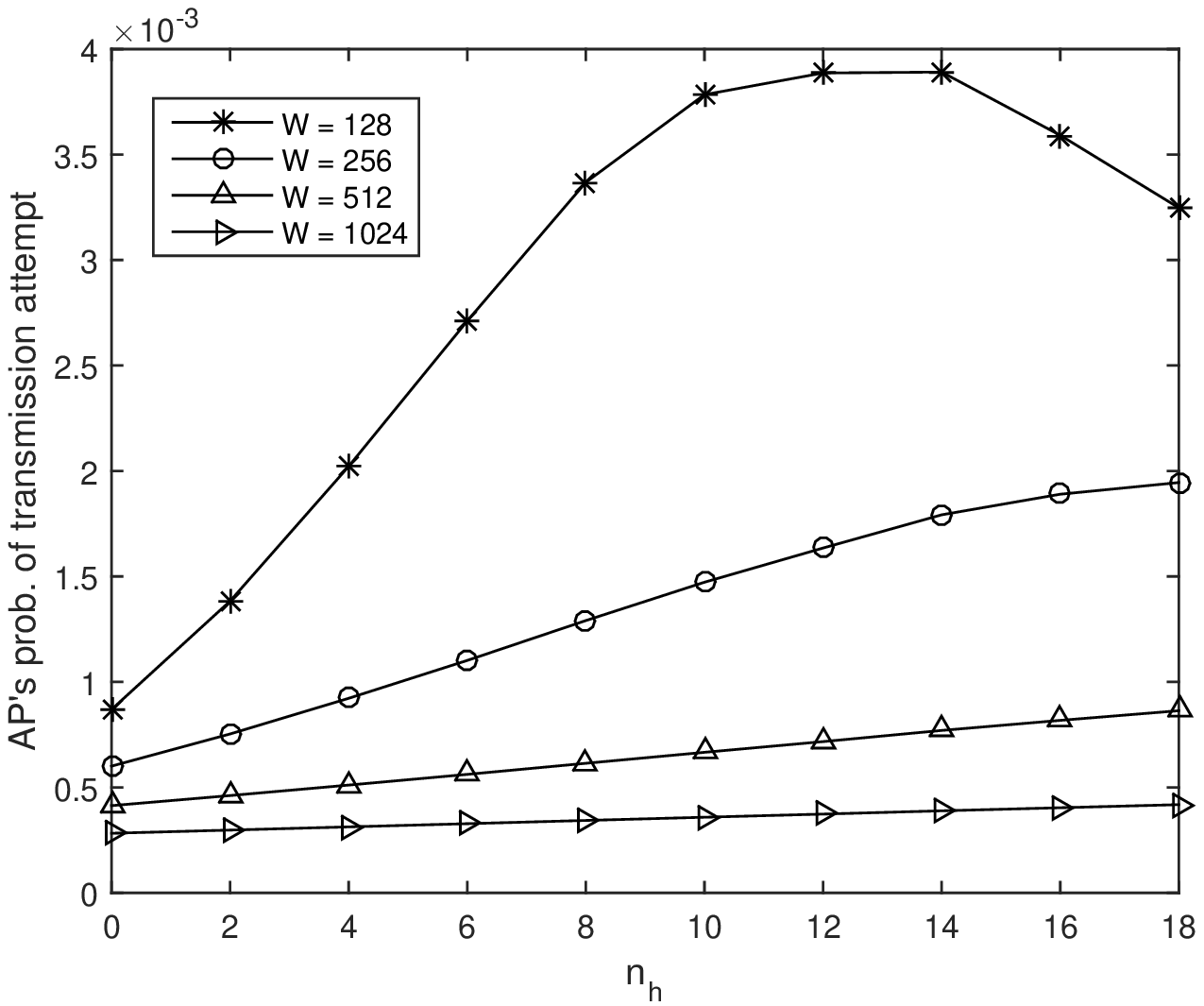} }}%
	\qquad
	\subfigure[$p^{ap}$]{{\includegraphics[width=0.38\textwidth]{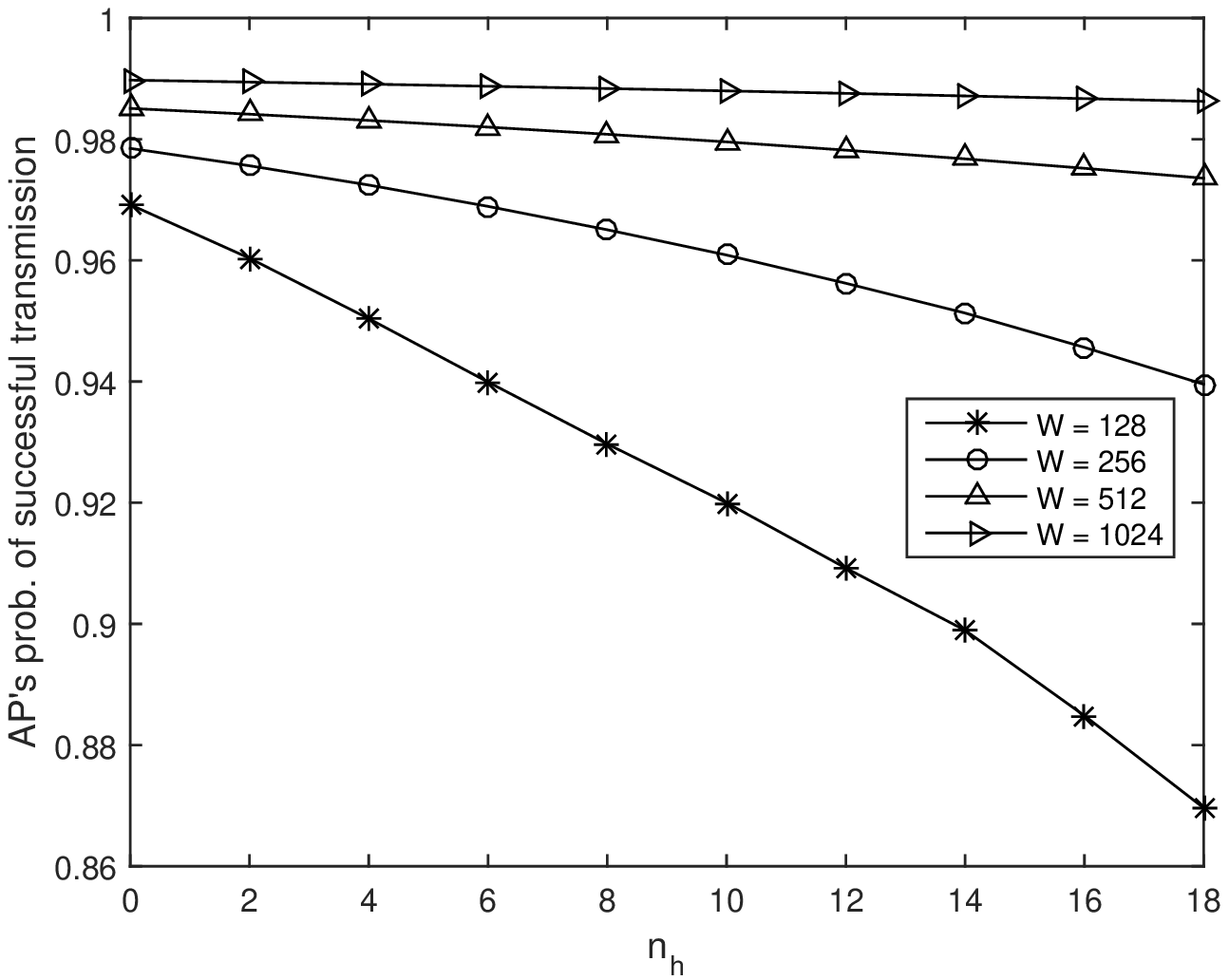} }}%
	\centering
	\caption{The effect of hidden terminals on $\omega^{ap}$ and $p^{ap}$}%
	\label{fig:example2}%
\end{figure*}

Figs. \ref{fig:example1} and \ref{fig:example2} show the behavior of parameters in (\ref{eq:gain_app}), namely $\omega$, $\omega^{ap}$, $p$ and $p^{ap}$, versus hidden terminals. For clients in Fig.~\ref{fig:example1}, $\omega$ increases gradually with hidden terminals due to an increase in channel idle time while $p$ decreases due to more collisions. In small windows such as $W=128$, $p$ converges to zero which shows the severity of hidden terminal problem and collisions. Though for the AP in Fig.~\ref{fig:example2}, $\omega^{ap}$ gradually increases while $p^{ap}$ experiences a rather slight decrease compared to the clients. The reason behind the increase in $\omega^{ap}$ is as the following. As the number of hidden terminals increases, collisions becomes much more frequent among clients leading to several retries and hence longer backoff for clients. During the back off and retries, collisions occupy the channel with shorter time than successful transmissions (collisions are detected during the header time). This translates to more channel idle time for the AP which leads to its more frequent transmission attempts. More transmission attempts always lead to more collisions hence the slight decrease in $p^{ap}$.

Based on this reasoning and looking at (\ref{eq:gain_app}), in high hidden terminal scenarios and low window sizes, $\omega^{ap}p^{ap}$ increases while $\omega p$ sharply decreases. The increase of term $\omega^{ap}p^{ap}$ and decrease of $\omega p$ leads to an increase in $\tilde{G}$. This explains the behavior we saw in Fig.~\ref{fig:gNnh}.       

% plot gain with rate or frame size.

Overall, we can conclude that in high hidden terminal scenarios we can expect a higher FD gain, albeit the total decrease in FD throughput. Topologically speaking, in scenarios where clients are concentrated at the edge of the cell, FD gain is higher when small windows are adopted.

%\section{Discussion}

%\section{Future Work}
%\subsection{Extending the model}

% discuss how this model is extendable to Multi-Ap networks.
%\subsection{Lessons for the design of full-duplex MAC}
%PCF-like techniques might be more suitable

\section{Conclusions and Design Guidelines}
\label{sec:conc}
In this paper, we presented an analytical model of the performance of CSMA/CA based MAC protocol for single-cell FD wireless networks. Packet-level simulations were used to validate the formulations of the model, and we demonstrated that they accurately estimate the saturation throughput for both FD and HD networks in the presence of hidden terminals. We point out the following design guidelines based on our observations:
\begin{itemize}
\item FD achieves higher performance than HD using RTS/CTS mechanism in any network configuration. However, when using CSMA/CA, the improvement over HD is about 35-40\% in the best case. Thus, the tradeoffs between the hardware complexity and the observed gains must be analyzed before deployment.

\item APs can effectively use FD only if they have an HOL packet destined for the client node that wants to initiate an FD transfer. Thus, AP designs must include packet fetching by lookups within the MAC queue in real time. The contention window changes must be done in a network-size cognizant manner as there exists a clear optimum gain for a specific network size and choice of the contention window.   

\item In an environment with significantly higher hidden terminals (say, nodes at the network edge), FD networks with smaller contention window values (around 128) perform about 25\% better than HD, though the raw throughput of FD can be improved by increasing the contention window. 

\item The same contention window gives both the maximum FD gain and the highest FD throughput. Thus, optimizing the window size for any one will suffice from a network design viewpoint. 

\item For moderate contention window sizes (512-1024) and in the absence of queue lookups for mitigating HOL blocking, approx. 25 clients or more per AP lowers the FD gain to the scenario of pure HD. Thus, for larger hotspots, mutli-cell FD networks should be designed.  
\end{itemize}
% We have considered the fact that the AP can only respond to a client's transmission in FD mode if and only if its HOL packet is destined for that node. Therefore as the network becomes denser and with uniform traffic assumption, the overall probability of an FD transaction gets smaller leading to lower FD gain versus HD. To improve FD's performance, fast and efficient packet queueing techniques must be designed in order to quickly search a packet for the transmitting client, while maintaining the physical layer delay constraints. Though we have addressed specifically the concerns of CSMA/CA based FD communications, this technology can be used in other channel access mechanisms as well.

Our current model only considers single-cell networks. To extend the FD analysis to multi-cell networks, as discussed above for larger networks, additional interference scenarios specific to these networks must be considered. These include pairs of APs and clients that fully or partly interfere and cause collisions. We will study these models in our future work.

%Based on this, one could conclude that centrally scheduled networks where each node is synchronized with a central authority might benefit more from the FD than a random access network such as the one considered here.

%\ifCLASSOPTIONcompsoc
  \section*{Acknowledgments}
%\else
The authors would like to thank the anonymous reviewers for their very constructive comments. This work was supported in part by the U.S. Office of Naval Research under grant number N000141410192.
%\fi

%The authors would like to thank...The authors would like to thank...The authors would like to thank...The authors would like to thank...The authors would like to thank...The authors would like to thank...The authors would like to thank...The authors would like to thank...The authors would like to thank...The authors would like to thank...The authors would like to thank...The authors would like to thank...The authors would like to thank...

\ifCLASSOPTIONcaptionsoff
  \newpage
\fi

\bibliographystyle{IEEEtran}
\bibliography{info}

\begin{IEEEbiography}[{\includegraphics[width=1in,height=1.25in,clip,keepaspectratio]{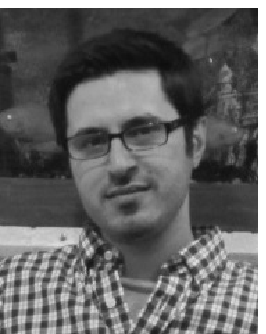}}]{Rahman Doost-Mohammady}
	received his B.Sc. in Computer Engineering in 2007 from Sharif University of Technology, Tehran, Iran, his M.Sc. Degree in Embedded Systems from Delft University of Technology in 2009, the Netherlands, and his PhD in Electrical and Computer Engineering from Northeastern University, Boston USA in 2014. Since January 2015, he has been a postdoctoral research associate at the ECE department, Northeastern University. His main research interests are wireless protocol design and performance analysis.   
\end{IEEEbiography}

\begin{IEEEbiography}[{\includegraphics[width=1in,height=1.25in,clip,keepaspectratio]{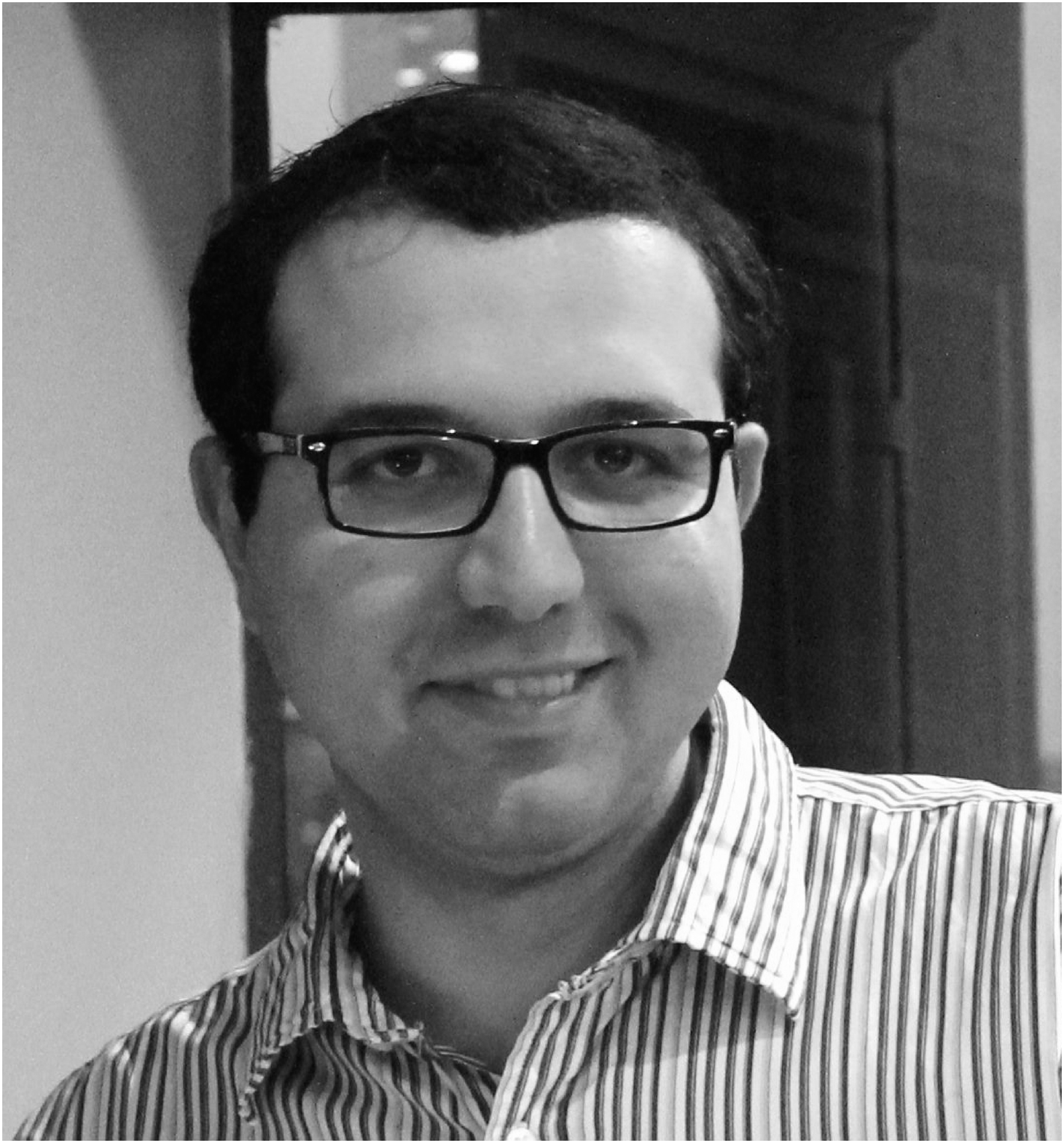}}]{M. Yousof Naderi}
	received the B.Sc. degree in computer engineering from Shahid Beheshti University (National University of Iran), Tehran, Iran, in 2008 and the M.Sc. degree with honors in communication and computer networks from Sharif University of Technology, Iran, in 2010. Currently, he is pursuing the Ph.D. degree in the Electrical and Computer Engineering Department at Northeastern University, Boston, MA, USA. His current research interests lie in the design and experimentation of novel communication protocols, algorithms, and analytical models, specialized in wireless energy harvesting networks, cognitive radio networks, multimedia sensor networks, and cyber-physical system.
\end{IEEEbiography}

%\newpage

\begin{IEEEbiography}[{\includegraphics[width=1in,height=1.25in,clip,keepaspectratio]{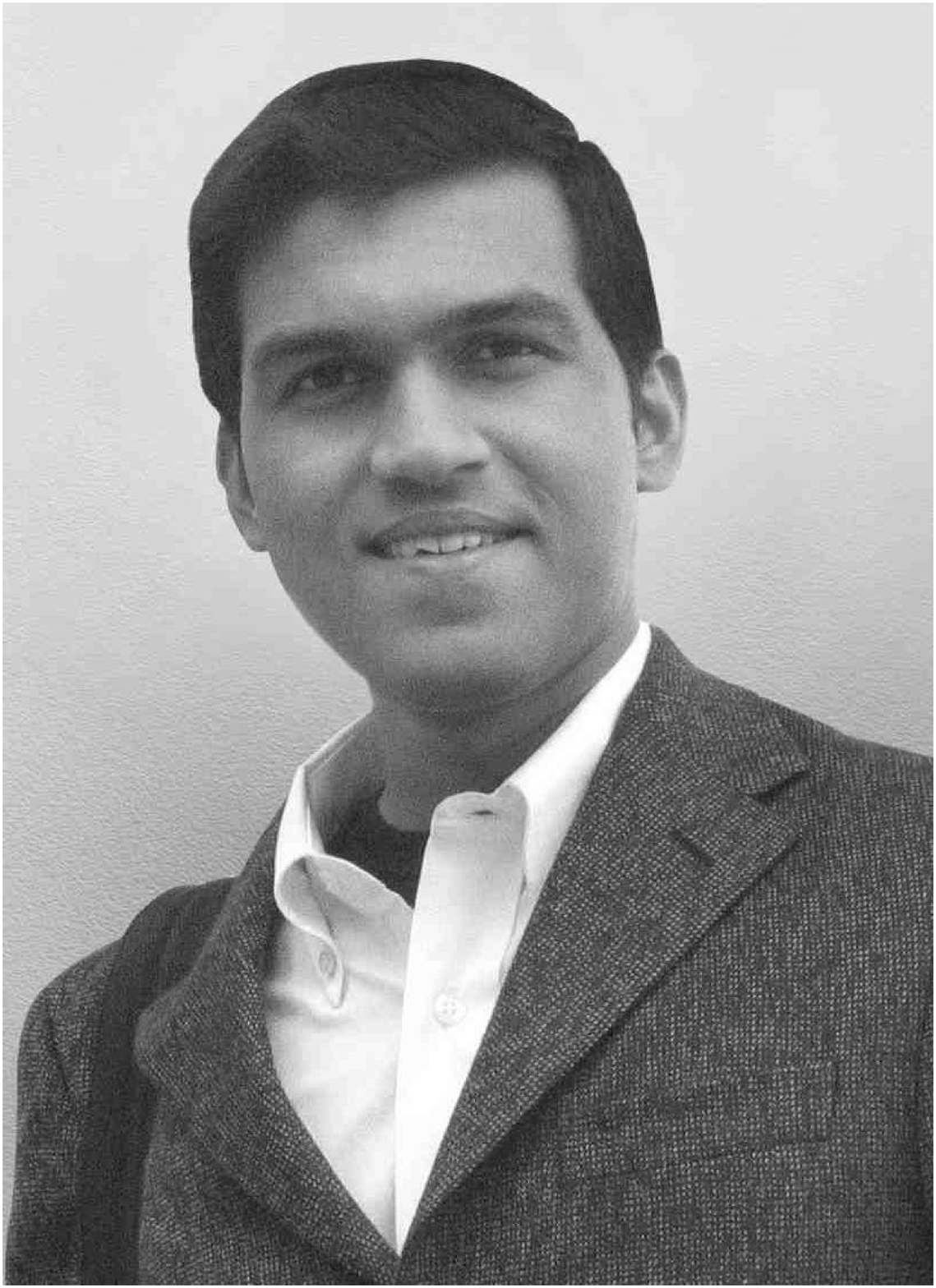}}]{Kaushik R. Chowdhury}
	received the B.E. degree in electronics engineering with distinction from VJTI, Mumbai University, India, in 2003, and the M.S. degree in computer science from the University of Cincinnati, Cincinnati, OH, in 2006, and the Ph.D. degree from the Georgia Institute of Technology, Atlanta, in 2009. His M.S. thesis was given the outstanding thesis award jointly by the Electrical and Computer Engineering and Computer Science Departments at the University of Cincinnati. He is Associate Professor in the Electrical and Computer Engineering Department at Northeastern University, Boston, MA. He currently serves on the editorial board of the Elsevier \textit{Ad Hoc Networks} and Elsevier \textit{Computer Communications} journals. His expertise and research interests lie in wireless cognitive radio ad hoc networks, energy harvesting, and intra-body communication. Dr. Chowdhury is the recipient of multiple best paper awards at the IEEE ICC conference.
\end{IEEEbiography}

\end{document}